\documentclass[a4paper,11pt]{article}
\pdfoutput=1
\usepackage[utf8]{inputenc}
\usepackage{amssymb}
\usepackage{amsmath}
\usepackage{amsthm}
\usepackage{amsfonts}
\usepackage[pdftex]{graphicx}
\usepackage{geometry}
\usepackage{braket}
\usepackage{enumerate}
\usepackage{comment}

\usepackage{subcaption}

\allowdisplaybreaks

\usepackage[table,usenames,dvipsnames]{xcolor}

\usepackage{setspace}
\usepackage[in]{fullpage}
\usepackage{hyperref}
\usepackage{array}
\usepackage{cancel}
\usepackage{wrapfig}
\usepackage[font=small,labelfont=bf]{caption}
\usepackage{pifont}

\usepackage[dvipsnames]{xcolor}
\usepackage{mathtools}

\usepackage{color}

\usepackage{tikz}
\usetikzlibrary{arrows,shapes}
\usetikzlibrary{trees}
\usetikzlibrary{matrix,arrows} % For commutative diagram
\usetikzlibrary{positioning}% For "above of=" commands
\usetikzlibrary{calc,through}% For coordinates
\usetikzlibrary{decorations.pathreplacing}% For curly braces
\usepackage{pgffor}	% For repeating patterns

\numberwithin{equation}{section}

\usepackage[english]{babel}

\usepackage{caption}

\usepackage[nottoc,notlot,notlof]{tocbibind}
\usepackage[nosort]{cite}   
\usepackage{color}

\usepackage{parskip}
\usepackage[T1]{fontenc}

\usepackage{lmodern}
\usepackage{sectsty}
\usepackage{yfonts}
\allsectionsfont{\boldmath}

\usepackage{physics}
\usepackage{slashed}
\usepackage{makeidx}
\usepackage[vcentermath]{youngtab}
\usepackage{amsthm}
\usepackage[scr=boondoxo]{mathalfa}

\hypersetup{
	colorlinks=true,
	linktoc=page,
	citecolor=blue,
	linkcolor=blue,
	urlcolor=blue} 

\usepackage{tikz}
\usetikzlibrary{arrows,shapes}
\usetikzlibrary{trees}
\usetikzlibrary{matrix,arrows} 				% For commutative diagram
\usetikzlibrary{positioning}				% For "above of=" commands
\usetikzlibrary{calc,through}				% For coordinates
\usetikzlibrary{decorations.pathreplacing}  % For curly braces
\usepackage{pgffor}							% For repeating patterns

\usetikzlibrary{decorations.pathmorphing}	% For Feynman Diagrams
\usetikzlibrary{decorations.markings}
\tikzset{
   vector/.style={decorate, decoration={snake, amplitude=1pt, segment length=6pt}, draw,double},
   vector2/.style={decorate, decoration={snake, amplitude=1pt, segment length=6pt}, draw},
	provector/.style={decorate, decoration={snake,amplitude=2.5pt}, draw},
	antivector/.style={decorate, decoration={snake,amplitude=-2.5pt}, draw},
    fermion/.style={draw=black, postaction={decorate},
        decoration={markings,mark=at position .55 with {\arrow[draw=black]{>}}}},
    fermionbar/.style={draw=black, postaction={decorate},
        decoration={markings,mark=at position .55 with {\arrow[draw=black]{<}}}},
    fermionnoarrow/.style={draw=black},
    gluon/.style={decorate, draw=black,
        decoration={coil,amplitude=4pt, segment length=5pt}},
    scalar/.style={dashed,draw=black, postaction={decorate},
        decoration={markings,mark=at position .55 with {\arrow[draw=black]{>}}}},
    scalarbar/.style={dashed,draw=black, postaction={decorate},
        decoration={markings,mark=at position .55 with {\arrow[draw=black]{<}}}},
    scalarnoarrow/.style={dashed,draw=black},
    electron/.style={draw=black, postaction={decorate},
        decoration={markings,mark=at position .55 with {\arrow[draw=black]{>}}}},
	bigvector/.style={decorate, decoration={snake,amplitude=4pt}, draw},
}
\tikzset{cross/.style={cross out, draw, 
         minimum size=2*(#1-\pgflinewidth), 
         inner sep=0pt, outer sep=0pt}}

\tikzstyle{block} = [draw, rectangle, 
    minimum height=3em, minimum width=6em]

\newcommand{\agl}[2]{\langle#1 #2 \rangle}
\newcommand{\sqr}[2]{\lbrack #1 #2 \rbrack}

\newcommand{\cA}{\mathcal{A}}
\newcommand{\cN}{\mathcal{N}}
\newcommand{\cL}{\mathcal{L}}
\newcommand{\cO}{\mathcal{O}}

\usepackage{float}
\usepackage{marvosym}
\usepackage{empheq}
\usepackage{bbold}

\usepackage{tikz}
\usetikzlibrary{shapes}
\usetikzlibrary{arrows}
\usetikzlibrary{positioning}
\usetikzlibrary{decorations.pathmorphing}
\usetikzlibrary{decorations.pathreplacing}
\usetikzlibrary{decorations.markings}

 \def\cA{\mathcal{A}}
 \def\uno{\mbox{1 \kern-.59em {\rm l}}}

\begin{document}

%%%% Title page %%%%

\begin{flushright}
	QMUL-PH-20-05\\
	SAGEX-20-04-E\\
	%HU-EP-18/25
\end{flushright}

\vspace{20pt} 

\begin{center}

	%{\Large \bf  Newton potential in $R^2$ gravity revisited   }  \\
	{\Large \bf  Eikonal phase matrix, deflection angle and time delay}  \\
	\vspace{0.3 cm} {\Large \bf  in effective field theories of gravity}

	\vspace{25pt}

	{\mbox {\sf  \!\!\!\!Manuel~Accettulli~Huber, Andreas~Brandhuber, Stefano~De~Angelis and 				Gabriele~Travaglini{\includegraphics[scale=0.05]{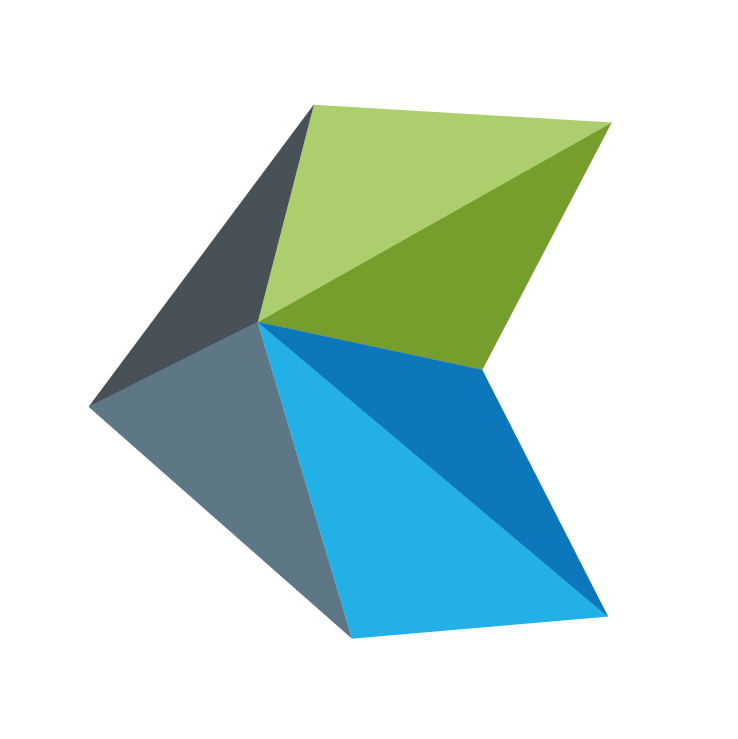}}
	}}
	\vspace{0.5cm}

	\begin{center}
		{\small \em
			Centre for Research in String Theory\\
			School of Physics and Astronomy\\
			Queen Mary University of London\\
			Mile End Road, London E1 4NS, United Kingdom
		}
	\end{center}

	%\vspace{-8pt}

	\vspace{40pt}  %was 40 

	{\bf Abstract}
\end{center}

\vspace{0.3cm}

\noindent
The eikonal approximation is an ideal tool to extract classical observables in gauge theory and gravity directly from scattering amplitudes. 
Here we  consider effective  theories of gravity where in addition to the Einstein-Hilbert term we include non-minimal couplings of the type $R^3$, $R^4$ and $FFR$.  In particular, we study the scattering of gravitons and photons of frequency $\omega$ off heavy scalars of mass $m$ in the limit  $m\gg \omega \gg |\vec{q}\,|$, where $\vec{q}$ is the momentum transfer. 
The presence of non-minimal couplings induces helicity-flip processes which survive the eikonal limit, thereby promoting the eikonal phase to  an eikonal phase matrix.
We obtain the latter from the relevant two-to-two  helicity amplitudes that we compute up to one-loop order,
and confirm that the leading-order terms in $\omega$ exponentiate \`{a} la Amati, Ciafaloni and Veneziano. From the eigenvalues of the 
eikonal phase matrix we then extract two physical observables, to 2PM order: the classical deflection angle and Shapiro
time delay/advance. 
Whenever the classical expectation of helicity conservation of the massless scattered particle is violated, {\it i.e.}~the eigenvalues of the eikonal matrix are non-degenerate, causality violation due to time advance is a generic possibility for small impact parameter.
We show that for graviton scattering in the $R^4$ and $FFR$ theories, time advance is circumvented if the couplings of these interactions satisfy certain positivity conditions, while it
is unavoidable for graviton scattering in the $R^3$ theory and photon scattering in the $FFR$ theory.
The scattering processes we consider mimic the deflection of photons and gravitons off spinless heavy objects such as black~holes.

\vfill
\hrulefill
\newline
\vspace{-1cm}
${\includegraphics[scale=0.05]{Sagex.jpeg}}$~\!\!{\tt\footnotesize\{m.accettullihuber, a.brandhuber, s.deangelis, g.travaglini\}@qmul.ac.uk}

\setcounter{page}{0}
\thispagestyle{empty}
\newpage

%%%%%%%%%%%%%%%%%% TABLE OF CONTENTS %%%%%%%%%%%%%%%%%%%%%%%%%%%%%%%%%

\setcounter{tocdepth}{4}
\hrule height 0.75pt
\tableofcontents
\vspace{0.8cm}
\hrule height 0.75pt
\vspace{1cm}
\setcounter{tocdepth}{2}

\newpage
%%%%%%%%%%%%%%%%%%%%%%%%%%%%%%%%%%%%%%%%%%%%%%%%%%%%%%%%%%%
\section{Introduction} 

\subsection{Overview} 
One of the  exciting applications of scattering amplitudes focuses on  the computation of classical observables in gauge theory and gravity such as deflection angles and time delay/advance, or effective Hamiltonians describing the dynamics of binary systems. Early results in this direction date  back to \cite{Iwasaki:1971iy},
where it was already noted that loop amplitudes contribute to classical processes, contradicting
the erroneous belief of {\it e.g.}~\cite{Feynman:1963ax}.
The intimate connection between loops  and classical physics was sharpened in \cite{Holstein:2004dn}, and had already been applied  in \cite{BjerrumBohr:2002kt} to obtain the classical and quantum  $\cO(G^2)$ corrections to Newton's potential, where $G$ is Newton's constant. In this approach, gravity is treated as an effective theory \cite{Donoghue:1994dn}, where one can make 
 predictions at low energy despite the non-renormalisability of the theory. 
More recently, a systematic approach employing scattering amplitudes in conjunction with unitarity \cite{Bern:1994zx,Bern:1994cg} was developed to compute classical quantities in gauge theory and gravity. 
Classical  \cite{Neill:2013wsa} and quantum  \cite{Neill:2013wsa,Bjerrum-Bohr:2013bxa} corrections to Newton's potential can be obtained from a two-to-two scattering amplitude of two massive scalars,  in particular narrowing down to terms 
  that have discontinuities in the channel corresponding to the momentum transfer $\vec{q}$ of the  process  
  \cite{Donoghue:1994dn, Holstein:2004dn}. An additional simplification stems from the fact that in the unitarity-based  calculation the cuts  can be kept in four dimensions, as discrepancies with $D$-dimensional results only give   rise to   analytic terms, at least  at one loop.  
 Unitarity was also applied in \cite{Bjerrum-Bohr:2014zsa, Bjerrum-Bohr:2016hpa, Bai:2016ivl, Chi:2019owc} to compute the deflection angle for light or for gravitons passing by a heavy mass, a quantity that  has the advantage of being   gauge invariant. 
We also mention some of the efforts leading to the computation of the effective (Newton)  potential  at  second 
\cite{Damour:1985mt,Gilmore:2008gq}, third \cite{Damour:2001bu, Blanchet:2003gy, Itoh:2003fy,Foffa:2011ub}, fourth \cite{Jaranowski:2012eb, Damour:2014jta,Galley:2015kus,Damour:2015isa, Damour:2016abl, Bernard:2015njp, Bernard:2016wrg, Foffa:2012rn, Foffa:2016rgu,Porto:2017dgs,Porto:2017shd,Foffa:2019yfl,Blumlein:2020pog}, fifth \cite{Foffa:2019hrb,Blumlein:2019zku}  and sixth \cite{Blumlein:2020znm} post-Newtonian order, following the landmark  computation at first post-Newtonian order \cite{Einstein:1938yz}. Comprehensive reviews on this topic from different perspective can be found in \cite{Foffa:2013qca,Rothstein:2014sra,Porto:2016pyg,Levi:2018nxp}.
In the post-Minkowskian framework, which is natural in the context of amplitudes, the current state of the art is at 3PM order  \cite{Bern:2019nnu,Bern:2019crd}, a result 
which was recently confirmed in \cite{Blumlein:2020znm,Cheung:2020gyp}.
Note also the effective one-body approach of \cite{Buonanno:1998gg},  recently extended to incorporate the first and second post-Minkowskian corrections in \cite{Damour:2016gwp,Damour:2017zjx}, respectively.
For other interesting approaches to extract classical observables in general relativity from amplitudes see \cite{Luna:2016due, Cachazo:2017jef, Bjerrum-Bohr:2018xdl, Plefka:2018dpa, Cheung:2018wkq, Kosower:2018adc, Guevara:2018wpp, Bautista:2019tdr, KoemansCollado:2019ggb, Bjerrum-Bohr:2019kec, Kalin:2019rwq}.

\subsection{Gravity with higher-derivative couplings}

Much attention has been  devoted  to the study of effective theories of gravity   obtained by adding 
 higher-derivative interactions to the 
Einstein-Hilbert (EH) action. In particular, efforts have been made  in  \cite{Endlich:2017tqa,Sennett:2019bpc} to confront such modifications with gravitational wave observations.  It   was also noted in \cite{Endlich:2017tqa} that for these effects to be measurable by experiments such as LIGO the cutoff of the effective theory must not be  much larger than  $\cO({\rm km}^{-1})$. 
In \cite{Brandhuber:2019qpg} we initiated a study of the effects that these higher-derivative terms
have on the Newtonian potential and deflection angle.
% in the framework of effective theory.
In this paper we sharpen this study by rooting it in the eikonal approximation  --  specifically, applying  it     to  three types of terms,     denoted schematically as $R^3$, $R^4$ and $FFR$, for which we  compute the corresponding corrections to the deflection angle and  time delay/advance.
More in detail, the particular  action  we consider for the graviton, photon and a massive scalar has the  form: 
\begin{equation}
\begin{split}
\label{action}
S = \int\!\dd[4] x\sqrt{-g} \,  \bigg[ &-\frac{2}{\kappa^2}R   \, - \, \frac{1}{4} F^{\mu \nu} F_{\mu\nu}\, +\, \frac{1}{2} (D_{\mu} \phi)( D^{\mu}\phi) - \frac{1}{2} m^2 \phi^2\\
&- \,  {2\over \kappa^2} \bigg( {\alpha^{\prime \, 2} \over 48} \, I_1 + {\alpha^{\prime \, 2} \over 24} \, G_3 \bigg)\, + \, \frac{2}{\kappa^2} \mathcal{L}_8 \, - \, \dfrac{\alpha_{\gamma}}{8} \, F^{\mu\nu} F^{\rho \sigma} R_{\mu \nu \rho \sigma}  \, \bigg] \, , 
\end{split}
\end{equation}
where 
\begin{align}
I_1 := {R^{\alpha \beta}}_{\mu \nu} {R^{\mu \nu}}_{\rho \sigma} {R^{\rho \sigma}}_{\alpha \beta}\ , \qquad 
G_3 := I_1 - 2 {R^{\mu \nu \alpha}}_\beta {R^{\beta \gamma}}_{\nu \sigma} {R^\sigma}_{\mu \gamma \alpha}\  , 
\end{align}
while  
\begin{equation}
    \label{L8}
    \cL_{8} \,=\, \beta_{1}\, \mathcal{C}^{\, 2}\, +\, \beta_{2}\, \mathcal{C}\, \widetilde{\mathcal{C}}\, +\, \beta_{3}\, \widetilde{\mathcal{C}}^{\, 2}\ ,
\end{equation}
where
\begin{equation}
    \label{eq:Cs}
    \mathcal{C} \,:=\, R_{\mu \nu \rho \sigma}\, R^{\mu \nu \rho \sigma}\ , \hspace{2cm} \widetilde{\mathcal{C}} \,:=\dfrac{1}{2}\, R_{\mu \nu \alpha \beta}\, \epsilon^{\alpha \beta}\,_{\gamma \delta}\, R^{\gamma \delta \mu \nu}\ .
\end{equation}
A few comments on the various couplings in \eqref{action}  are in order here.

%{\bf 1.}~
First, there are two types of $R^3$ terms, denoted as $I_1$ and $G_3$ above. Such terms arise naturally in the  low-energy effective description of bosonic string theory. Their effects on gravitational scattering of different matter fields have been discussed recently in \cite{Brandhuber:2019qpg, Emond:2019crr}; specifically for the  scattering of two massive scalars, both independent structures $I_1$ and $G_3$ were found to contribute. On the other hand, for the helicity-preserving deflection of massless particles of spin 0, 1 and 2, it was  shown in \cite{Brandhuber:2019qpg} that the  $G_3$ interaction has no effect.  
Additional interesting features  about the $I_1$ and $G_3$ couplings are that  $I_1$ is the only coupling  that contributes to pure graviton scattering up to four points \cite{doi:10.1063/1.523128, Broedel:2012rc}
and is the two-loop counterterm in pure gravity \cite{Goroff:1985sz}, while $G_3$ is a topological term in six dimensions. In the following we will be concerned with  (helicity-preserving and flipping) scattering of massless gravitons in the background produced by a massive scalar, in which case only the $I_1$ structure contributes, hence we will refer to it simply as the $R^3$ term, since no confusion can arise.  
Note that in the case of photons there is no $R^3$ contribution to the helicity-flipping process.

%{\bf 2.}~
The second interaction we study is of the type $R^4$. In principle there are  26 independent   parity-even quartic contractions of the Riemann tensor  \cite{Fulling:1992vm},  but only the seven which do not contain the Ricci scalar or tensor survive on shell  in arbitrary dimensions, as can also be seen using field redefinitions \cite{Tseytlin:1986zz, Metsaev:1987zx, Huber:2019ugz}. In four dimensions these reduce to two independent parity-even structures \cite{Endlich:2017tqa,Dunbar:2017szm}, plus one parity-odd structure \cite{Sennett:2019bpc}, as shown in \eqref{L8}.% 
\footnote{A general approach to find a  complete, non-redundant operator basis of  dimension six and eight for 
the effective Standard Model including gravity has been given recently in \cite{Ruhdorfer:2019qmk}
using the Hilbert series method.}
In agreement with
\cite{Dunbar:2017szm} we find that  these interactions induce  the following four-point graviton amplitudes: those  with all-equal helicities, and the amplitude  with two positive- and two negative-helicity gravitons (the MHV configuration). If $\beta_2$ in \eqref{L8} is non-vanishing, then the all-plus and all-minus graviton amplitudes are  independent. 
We also note that a particular contraction of four Riemann tensors appears %for example 
in type-II superstring theories where it is the first  higher-derivative curvature correction to the EH theory, and can
be determined from four-graviton scattering \cite{Gross:1986iv}.

The  third interaction we consider is an $FFR$ term, where $F$ is the electromagnetic field strength. It is known to arise in string theory as well as from integrating out massive, charged electrons in the case of electrodynamics coupled to gravity, as discussed  in \cite{Berends:1975ah,Drummond:1979pp}, and considered more recently in \cite{Hollowood:2007ku,Goon:2016une}.

We have also introduced in the action a minimally coupled  massive scalar to represent a black hole%
\footnote{In order to describe charged black holes the real scalar in \eqref{action} should be replaced by an electrically charged complex scalar.}.
Note that in \eqref{action} we have excluded terms quadratic in the curvatures since from an effective field theory/on-shell point of view they have no effect to any order in four dimensions, as shown recently in \cite{Huber:2019ugz}.

\subsection{Physical observables from the eikonal phase matrix}

We now come to the computation of the physical observables of interest  -- these   are the classical deflection angle 
and the time delay/advance \cite{Shapiro:1964uw} experienced by  massless gravitons and photons when they scatter off a (possibly charged) massive scalar. 
A method ideally suited for obtaining classical observables directly from amplitudes, without passing through
intermediate, unphysical quantities, is the eikonal  \cite{Cheng:1969eh,Sucher:1969,Abarbanel:1969ek,Amati:1987wq,Amati:1990xe,Kabat:1992tb}. 
In this approach the relevant amplitudes are evaluated in an approximation where the momentum transfer $|\vec{q}\, |$ is taken to be much smaller than both the mass $m$  of the heavy scalar and the energy  $\omega$ of the massless particle, or more precisely taking $m \gg \omega \gg |\vec{q}\, | $. Crucial for this is a convenient parameterisation of spinor helicity variables for the massless particles in the eikonal limit.
The amplitudes thus obtained are then transformed to impact parameter space via a two-dimensional Fourier transform. In this  space the amplitudes are expected to exponentiate into an eikonal phase, from which one can extract directly the classical (and, if desired, quantum) deflection angle and time advance/delay.  
Recent applications of this method to this type of problem include \cite{Akhoury:2013yua} for the deflection angle of massless scalars up to 2PM, \cite{Bjerrum-Bohr:2016hpa} for photons and  fermions up to 2PM order, and up to 3PM order in \cite{Bern:2020gjj, Bern:2020buy, Parra-Martinez:2020dzs}. 
We also note that  \cite{Bjerrum-Bohr:2016hpa}  showed the equivalence of the eikonal method  and the formalism based on the computation of an intermediate potential/Hamiltonian used for instance in \cite{ Bjerrum-Bohr:2014zsa, Bjerrum-Bohr:2016hpa, Bai:2016ivl, Chi:2019owc,Brandhuber:2019qpg,Emond:2019crr}.

An important point we wish to make is that in our case, because helicity-preserving as well as helicity-violating processes contribute, the eikonal {\it phase} is promoted to an eikonal {\it phase matrix} in the space of helicities of the external massless particles,  with $(+-)$ and $(-+)$ being the diagonal entries associated to no-flip scattering (in a convention where all particles' momenta are outgoing), while $(++)$ and $(--)$ are the off-diagonal entries, with helicity flip.
The associated mixing problem  has to be resolved in order to obtain the  physical quantities of interest. Whenever the two eigenvalues of the eikonal phase matrix are distinct, a possible violation of causality at small impact parameter arises, as noticed already at tree level in \cite{Camanho:2014apa}. 
See also \cite{Hollowood:2007ku,Goon:2016une,Hollowood:2007kt,Hollowood:2015elj,Benakli:2015qlh,Hollowood:2016ryc,deRham:2019ctd,1798363} for further discussions an  resolutions of this issue  in UV-complete theories, \cite{Amati:1987wq,Camanho:2014apa,Amati:1987uf,DAppollonio:2013mgj} for 
earlier appearances of  the eikonal operator and \cite{Berends:1975ah,Drummond:1979pp} for related discussions involving helicity flip and no-flip amplitudes. 

\subsection{Summary of the paper}

We now summarise our results. We have computed the graviton deflection angle and time delay/advance for the three interactions $R^3$, $R^4$ and $FFR$, and in addition the photon deflection and time delay induced by the $FFR$ interaction. 
The single most important qualitative difference  with the EH  theory is  that the propagation and speed  of the massless particle  acquire a dependence on its polarisation in all  cases except the graviton propagation in the presence of the $FFR$ interaction. 
This  generically leads to a time advance at small impact parameter $b$. Interestingly, in the case of graviton scattering due to $R^4$ and $FFR$, causality violation can be avoided
if the coefficients of the interactions obey certain positivity constraints which, for $R^4$, are in precise agreement   with those of \cite{Gruzinov:2006ie, Bellazzini:2015cra}. For the $R^3$ interaction our results are fully consistent with the tree-level findings of \cite{Camanho:2014apa}, extending them to one loop. Note that while we used a massive scalar, \cite{Camanho:2014apa} used a coherent state to set up the background in which the graviton is deflected. Similarly, the $FFR$ interaction induces superluminal propagation of photons.

An important  point is the dependence of the eikonal $S$-matrix $S_{\rm eik}$ on the energy $\omega$ of the scattered massless particle. In the EH theory, $S_{\rm eik}$ is expected to take the form $S_{\rm eik} = e^{i (\delta_0+ \delta_1+ \cdots ) }$, where the subscript $L$ denotes the loop order of $\delta_L$. 
For the leading eikonal $\delta_0$, this was proven for our kinematic set-up in \cite{Akhoury:2013yua}, and it is   generally expected that the  $\delta_L$ are linear in $\omega$, although we are not aware of an all-order proof. Both for $R^3$ and $FFR$ our results are perfectly aligned with this expectation up to 2PM order, resulting in an  $\omega$-independent deflection angle and time advance/delay. A novelty arises for  $R^4$  where the corresponding eikonal phase (matrix) scales as $\omega^3$ with the graviton frequency.

The rest of the paper is organised as follows. In Section \ref{sec:2} we discuss our kinematic set-up and provide explicit expressions for the spinor helicity variables associated to each massless particle in the eikonal limit. We then discuss some general aspects of the eikonal approximation,  in particular the extraction of the phases $\delta_L$ from the loop scattering amplitudes. We highlight  consistency conditions relating amplitudes in impact parameter space at different loop orders and powers of $\omega$ which will then be explicitly checked in all cases considered. We also quote here the relevant formulae to derive the particle deflection angle and time advance/delay. 

Section \ref{sec:3} contains the computations of all  tree-level and one-loop amplitudes relevant for our analysis. As a warm up we consider the EH theory, where we re-discuss the graviton deflection computation of \cite{Chi:2019owc}; while it is conventionally assumed that in the classical picture the helicity of the scattered massless projectile is unchanged, we show explicitly that this is the case {\it in the eikonal limit}: the flipped-helicity amplitudes are  non-zero both at tree level and one loop, but are subleading once the eikonal limit is taken, resulting in a diagonal eikonal phase matrix. We then move on to present the relevant four-point two-scalar two-graviton amplitudes with and without helicity flip in the case of $R^3$, $R^4$ and $FFR$, as well as the two-scalar two-photon amplitudes for the $FFR$ case, all at tree and one-loop level. Some of these amplitudes have been calculated here for the first time.
%(and we summarise them in Appendix \ref{app:C} for the reader's convenience).
While at tree level we present exact expressions, at one loop we work in the eikonal approximation and we only consider cuts  in the $q^2$-channel which produce  non-analytic terms
arising from the long-range propagation of two  massless particles, following the approach initiated in  \cite{BjerrumBohr:2002kt,Neill:2013wsa,Bjerrum-Bohr:2013bxa,Khriplovich:2004cx,Iwasaki:1971vb}. Although in the subsequent section we focus only on classical contributions, arising from triangle integral functions with one internal mass, we quote  complete answers for the amplitudes up to one loop including box (needed to check the exponentiation of the tree-level phase matrix) and bubble integrals (generating $\cO(\hbar)$ corrections to the physical observables). 

Section \ref{sec:4} is dedicated to the computation of the leading and subleading eikonal matrices $\delta_0$ and $\delta_1$, from which we will then obtain the $\cO(G)$ and $\cO(G^2)$ corrections to the deflection angle and time advance/delay for the four cases considered  -- scattering of gravitons in the presence of $R^3$, $R^4$ and $FFR$ terms, and scattering of photons induced by the $FFR$ interaction. We also show the case of  graviton scattering in EH to set the scene for the more complicated examples discussed later. In all cases we check the exponentiation of the tree-level eikonal phase matrix explicitly, providing important consistency checks of our calculations.  Our main results  are given in \eqref{main1}, \eqref{main3}, \eqref{RN1}, and in \eqref{main2}, \eqref{main4}, \eqref{RN2}, for the graviton deflection angle and time advance/delay in the $R^3$, $R^4$ and $FFR$ cases, while the photon deflection and time advance/delay in the $FFR$ theory are given in  \eqref{main7} and \eqref{main8}, respectively. 

A few appendices complete the paper, where we present relevant integrals, the Feynman rules used for some of our new computations, a list of tree amplitudes, and a derivation of the four-point graviton amplitudes in $R^4$ only based on little-group and dimensional analysis considerations.

\section{From amplitudes  to the deflection angle  and time delay via the eikonal}

\label{sec:2}

In this section we first give a precise definition of the eikonal limit providing an explicit parametrisation for all the momenta and spinor-helicity variables we  need. We then  briefly review the connection between amplitudes in the eikonal limit  (Fourier-transformed to impact parameter space) and the  eikonal phase matrix,  the deflection angle and the time delay.

\subsection{Kinematics of the scattering}\label{sec:kinematics}

We begin by  describing the kinematics of the scattering processes we  consider. We denote by  $p_1$ and $p_2$  the four-momenta of the incoming and outgoing scalars, respectively, with  $m$ being their common mass, while the momenta of the incoming and outgoing  massless particles (gravitons or photons) are $p_4$ and $p_3$. We will  work in the centre of mass frame, with the following parameterisation: 

\begin{equation}\label{kinematics}
\begin{array}{lr}

\begin{tikzpicture}[scale=15,baseline={([yshift=-1mm]centro.base)}]
\def\x{0}
\def\y{0}

\node at (0+\x,0+\y) (centro) {};
\node at (-3pt+\x,-3pt+\y) (uno) {$p_1$};
\node at (-3pt+\x,3pt+\y) (due) {$p_2$};
\node at (3pt+\x,3pt+\y) (tre) {$p_3^{h_3}$};
\node at (3pt+\x,-3pt+\y) (quat) {$p_4^{h_4}$};

\draw [thick] (uno) -- (centro);
\draw [thick] (due) -- (centro);
\draw [vector,double] (tre) -- (centro);
\draw [vector,double] (quat) -- (centro);

\draw [->] (-2.8pt+\x,-2pt+\y) -- (-1.8pt+\x,-1pt+\y); 
\draw [->] (2.8pt+\x,-2pt+\y) -- (1.8pt+\x,-1pt+\y); 
\draw [->] (-1.8pt+\x,1pt+\y) -- (-2.8pt+\x,2pt+\y); 
\draw [->] (1.8pt+\x,1pt+\y) -- (2.8pt+\x,2pt+\y); 

%\node at (0+\x,0+\y) [draw, fill=gray!90!black, circle, inner sep=10pt] {};

\shade [shading=radial] (centro) circle (1.5pt);

\end{tikzpicture}
&\hspace{2cm}
\begin{aligned}
p_4^\mu & = - (E_4,-\vec{p}+\vec{q}/2)\, ,  \\
 p_1^\mu & =  -(E_1,\vec{p}-\vec{q}/2) \, , \\
p_2^\mu & =  (E_2,\vec{p}+\vec{q}/2) \, ,  \\
p_3^\mu & =  (E_3,-\vec{p}-\vec{q}/2)\ .
\end{aligned}

\end{array}
\end{equation}
In our conventions we take all momenta to be outgoing, hence the minus signs in the expressions of  $p_1$ and $p_4$ since particles 1 and 4 are incoming. 
We also have 
\begin{align}
\begin{split}
\label{enbend}
E_1&=E_2 =\sqrt{m^2+\vec{p}^{\, \, 2}+\vec{q}^{\, \, 2}/4}\, , 
\\
E_3&=E_4=\sqrt{\vec{p}^{\,\,   2}+\vec{q}^{\, \, 2}/4}\,  := \, \omega
\ ,
\end{split}
\end{align}
 where  $\vec{p} \, \cdot \, \vec{q}=0$ due to momentum conservation.  Hence $\vec{q}$\,
lives in a two-dimensional space orthogonal to $\vec{p}$.
 In this paper  we  define  the  Mandelstam variables as
 \begin{align}
\label{mandel}
s:=(p_1+p_2)^2 = -\vec{q}^{\, \, 2}   , \ \  \ t:=(p_1+p_4)^2 = (E_1+E_4)^2  ,  
 \ \ \ u:=(p_1+p_3)^2  , 
 \end{align}
with $s+t+u = 2 m^2$.
In this notation the spacelike momentum transfer squared is given by $s$, while $t$ denotes the centre of mass
energy squared, and $\omega$ is the energy of the scattered massless particle.

In the above  parameterisation, the kinematic limit we are interested  is  
\begin{align}
\begin{split}
\label{bendlim}
m\gg \omega \gg |\vec{q}\, |
\ ,
\end{split}
\end{align}
which implies for the Mandelstam variables
\begin{equation}
t \simeq m^2 + 2m \omega \>, \hspace{1cm} ut - m^4 \simeq - (2 m \omega)^2 \>,
\end{equation}
and for the energies of the massless particles
\begin{align}
\begin{split}
E_3=E_4 := \omega \simeq |\vec{p}\, |\left(1 + {\vec{q}^{\, \, 2} \over 8 \,\vec{p}^{\, \, 2}}\right)
\ .
\end{split}
\end{align}
For definiteness we choose 
%the massless particle to move mainly along the $\hat{z}$ axis 
$\vec{p} = |\vec{p}\,|\, \hat{z}$ with $|\vec{p}\, |\gg |\vec{q}\, |$, as implied by \eqref{bendlim}.
In this approximation we can write the four-momentum $p_3$ of the massless particle in spinor notation as 
\begin{align}
\label{spainors}
\begin{split}
p_3 = \begin{pmatrix} 
\dfrac{\vec{q}^{\, \, 2}}{  8  |\vec{p}\, |}  & - \dfrac{\bar{q}}{2}
\vspace{0.3cm} \\ 
- \dfrac{{q}}{2}  & 2  |\vec{p}\, |
\end{pmatrix} \, , 
\end{split}
\end{align}
with $q:= q_1 + i q_2$ and $\bar{q}:= q_1 - i q_2$.  One can then find an explicit parameterisation for the spinors associated to the null momenta  $p_i=\lambda_i \tilde{\lambda}_i$, $i=3,4$, with 
the result 
\begin{align}
\begin{split}
\lambda_3 &= \sqrt{ 2  |\vec{p}\, |} 
\begin{pmatrix} 
 - \dfrac{\bar{q}}{ 4   |\vec{p}\, |}
\vspace{0.3cm} \\ 
1
\end{pmatrix} \ , \qquad 
\ \, \tilde{\lambda}_3 = \sqrt{ 2  |\vec{p}\, |} 
\begin{pmatrix} 
 - \dfrac{{q}}{ 4   |\vec{p}\, |} & 
\hspace{0.3cm}  
1
\end{pmatrix} \ , 
\\
\lambda_4 &=i \sqrt{ 2  |\vec{p}\, |} 
\begin{pmatrix} 
  \dfrac{\bar{q}}{ 4   |\vec{p}\, |}
\vspace{0.3cm} \\ 
1
\end{pmatrix} \ , \qquad \quad
\tilde{\lambda}_4 = i\sqrt{ 2  |\vec{p}\, |} 
\begin{pmatrix} 
  \dfrac{{q}}{ 4   |\vec{p}\, |} & 
\hspace{0.3cm}  
1
\end{pmatrix} \ .
\end{split}
\end{align}
Note the extra factors of $i$ due to the negative energy-component of $p_4$ corresponding to an incoming particle.

\subsection{Eikonal phase, deflection angle and time delay}

In this section we briefly review relevant aspects of the eikonal approximation and 
the eikonal phase matrix which allows for an efficient extraction
of the deflection angle and time delay/advance from scattering amplitudes. This topic was intensively studied in the context of gravity and string theory in the nineties~\cite{Amati:1990xe,Kabat:1992tb}; for
related recent work see also \cite{DiVecchia:2019myk,DiVecchia:2019kta,KoemansCollado:2019ggb} and references therein.

First, we introduce the amplitude in impact parameter space $\widetilde{\cA}$. This is defined as a  Fourier transform of the amplitude $\mathcal{A}$ with respect to the  momentum transfer  $\vec{q}$, 
\begin{align}
\label{amptraimp}
\widetilde{\cA} (\vec{b}\,) \ := \ {1\over 4 m \,\omega} \int\!{d^{D-2} q\over (2 \pi)^{D-2}} \, e^{i \vec{q} \cdot \vec{b}} \ \cA (\vec{q}\,)\ , 
\end{align}
where $\vec{b}$ is the impact parameter, and the number of dimensions will eventually be set  to $D=4-2\epsilon$.

In the eikonal approximation the gravitational $S$-matrix can be written in the form \cite{Amati:1990xe, Akhoury:2013yua}
\begin{align}
\label{eq:Seikonal0}
S_{\rm eik} = e^{i (\delta_0+ \delta_1+ \cdots ) } \ , 
\end{align} 
where $\delta_0$ is the leading eikonal phase, which is  $\mathcal{O}(G)$, $\delta_1$ the first subleading correction, of  $\mathcal{O}(G^2)$,  and the dots represent  subsubleading contributions. Alternatively, one can write the $S$-matrix in impact parameter space as
\begin{align}
\label{eq:Seikonal}
S_{\rm eik} \ = \ 1 +  \widetilde{\cA}^{(0)}_{\omega} +  \widetilde{\cA}^{(1)}_{\omega^2} + \widetilde{\cA}^{(1)}_{\omega} + 
 \widetilde{\cA}^{(2)}_{\omega^3} +\widetilde{\cA}^{(2)}_{\omega^2}+\widetilde{\cA}^{(2)}_{\omega}+\cdots \ , 
\end{align}
%\textcolor{red}{(Manuel and Stefano) Note that for $R^4$ this is not true any more.}
where the superscript indicates the loop order $L$ and the subscript the power in the energy $
\omega$ of the massless particle. That the maximal power of $\omega$ at a given loop order is $L+1$ is a well-established fact in (super)gravity and we will see below that the $R^3$
corrections do not  alter this expectation. However, we also find that the $R^4$ corrections lead to higher powers of 
$\omega$ starting at one loop, which is not surprising since  higher-derivative corrections worsen the high-energy behaviour. In the  effective field theory approach we  adopt, we are not really interested in high-energy physics (or high-energy completions of the theory) -- we use the eikonal approximation as an efficient and elegant  tool to extract deflection angles and time delay/advances without passing through  the computation of  non gauge-invariant intermediate quantities such as effective  potentials or Hamiltonians. Nevertheless  it would interesting to check if in the $R^4$ case unitarity can be restored as well through exponentiation.

Equating \eqref{eq:Seikonal0} with \eqref{eq:Seikonal} one gets 
\begin{align} 
\label{delta0}
\delta_0 &= \  -i \, \widetilde{\cA}^{(0)}_{\omega}\ ,\\[.2em]
\label{delta1}
\delta_1 &= \  -i\,  \widetilde{\cA}^{(1)}_{\omega}\ , 
\end{align}
as well as the condition 
\begin{align} 
- {(\delta_0)^2\over 2} \ = \  \widetilde{\cA}^{(1)}_{\omega^2}
\ , 
\end{align}
which implies the consistency condition 
\begin{align}
\label{exponentiation}
\widetilde{\cA}^{(1)}_{\omega^2} \ = \   {1\over 2} (\widetilde{\cA}^{(0)}_{\omega})^2
 \ . 
 \end{align}
Thus, the contribution to the one-loop amplitude that is  leading in $\omega$,  $\widetilde{\cA}^{(1)}_{\omega^2}$,  does  not provide any new information about the $S$-matrix. In general, it is only the term in $\widetilde{\cA}^{(L)}$ that is linear in $\omega$, $\widetilde{\cA}^{(L)}_{\omega}$, that provides new information entering $\delta_L$. We also note that \eqref{delta0}--\eqref{exponentiation} hold as  matrix equations. 

Note that a priori these statements are known to hold for EH gravity. The results in this paper show that
\eqref{exponentiation} also holds for the higher-derivative interactions discussed here at least up to one loop.
Of course the work of \cite{Amati:1990xe} on the eikonal limit of string  amplitudes gives reason to believe that the exponentiation will work for higher-derivative interactions to all orders.

Finally, the particle deflection angle can  be obtained from the eigenvalues $\delta^{(i)}$ of the eikonal phase matrix $\delta$. 
Using a saddle-point approximation \cite{Amati:1990xe, Bjerrum-Bohr:2016hpa,Bjerrum-Bohr:2017dxw} one finds, for small $\theta$, 
\begin{equation}
\label{rebend}
\theta^{(i)}  \ = \ \frac{1}{\omega} \frac{\partial}{\partial b} {\delta^{(i)}} \, , 
\end{equation}
where  $i$ runs over all  eigenvalues of $\delta$ and $b=|\vec{b}\,|$. 
For the time delay, we will use  instead \cite{Eisenbud:1948paa,Bohm,Wigner:1955zz}
 \begin{equation}\label{eq:timedelay}
t^{(i)} = \frac{\partial \delta^{(i)}}{\partial \omega}
\ .
\end{equation}

\section{The relevant scattering amplitudes}
\label{sec:3} 

In this section we compute the relevant  amplitudes needed to extract the  deflection  angle and time delay/advance induced  by the various interactions    in \eqref{action}. At tree level we will present exact expressions; at one loop  we only need to compute the part of the amplitude with  a discontinuity in the $s$-channel%
\footnote{We recall that  $s=-|\vec{q}\, |^2$ where  $\vec{q}$ is the  momentum exchange between the classical source and the graviton. } 
and we will write  the relevant expressions after expanding them in the eikonal approximation \eqref{bendlim}
-- this will be denoted in the following by the $\simeq$ symbol. 
A direct extraction of the classical part of the deflection  angle and time delay can be  performed  using triple cuts, and in an even more refined way using the holomorphic classical limit \cite{Guevara:2017csg}. We chose instead to compute the one-loop amplitudes through two-particle cuts, which also determine the quantum part of the amplitude. The latter, despite not being used in the present paper, becomes essential when considering the exponentiation in the eikonal limit at higher orders \cite{DiVecchia:2019kta}.

We will begin  our discussion with the simple case of EH gravity, quoting  from  \cite{Chi:2019owc} the relevant two-scalar two-graviton amplitude without helicity flip. We also compute the amplitude with helicity flip, and show that it does not contribute in the eikonal approximation, as correctly assumed in previous treatments. We will then move on to compute the relevant tree and one-loop amplitudes that are necessary in order to compute the  corrections induced by the $R^3$, $R^4$ and $FFR$ terms in \eqref{action}.

\begin{figure}[ht]
    \centering
    \includegraphics[scale=1]{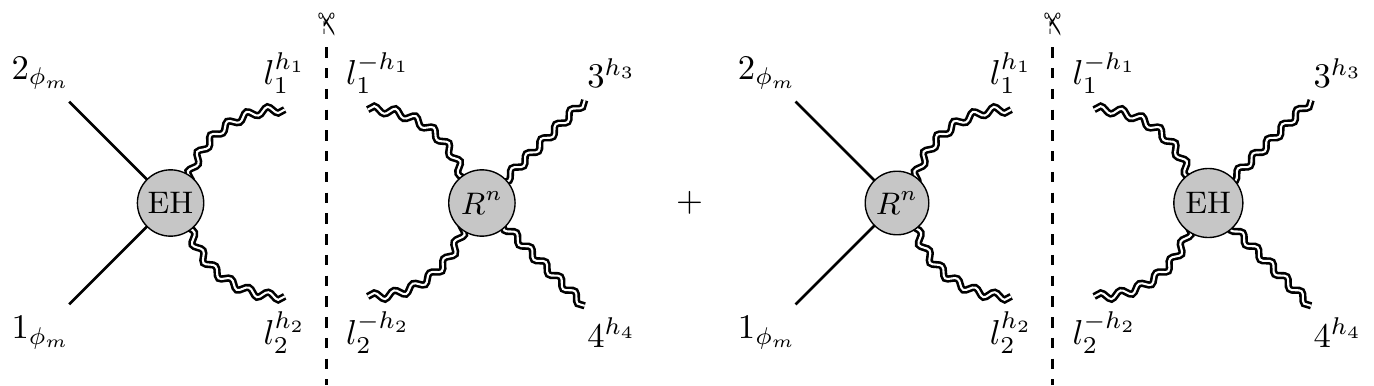}
    \caption{The two-particle cut diagrams for the $R^n$ interaction in the $s=-\vec{q}^{\, \, 2}$-channel. In our conventions external momenta are all outgoing and internal loop momenta flow from left to right in the diagram.}
        \label{fig:doublecut}
\end{figure}

The   two-particle cut diagrams relevant for the   $R^3$ and $R^4$ cases are shown in Figure~\ref{fig:doublecut}. The corrections induced by the  $FFR$ interaction need a separate analysis and we show the corresponding  diagrams  in Figures \ref{fig:doublecutFFRgraviton} and \ref{fig:doublecutFFRphoton}. 
For the case of the $R^n$ interaction both internal and external particles are gravitons, while in the case of $FFR$ we either have external gravitons and internal photons, or viceversa.

A comment is in order here. 
Focusing on the cuts relevant for $R^n$ depicted in Figure~\ref{fig:doublecut}, 
 the case  $h_3=h_4$ corresponds to the massless particle flipping helicity upon interacting with the scalar, whereas $h_3=-h_4$ corresponds to the helicity-preserving process, since in our conventions all external particles are  outgoing.
A simple way to take into account particle statistics is to sum over all values of the internal helicities $h_1$ and $h_2$ and 
divide the result by~2.%
\footnote{If the two particles are identical this introduces the correct Bose symmetry factor of $1/2$; if they are different this takes into account that the internal particles are not colour ordered, hence summing over two possible internal helicity assignments   would lead to double counting, compensated by the factor of $1/2$.}

\subsection{Four-point scalar/graviton scattering in EH  gravity}

The relevant tree-level amplitudes  in the EH case are the two-scalar/two-graviton amplitudes in the two helicity configurations for the gravitons:%
\footnote{See for instance \cite{Brandhuber:2019qpg,Nandan:2018ody}.}

\begin{equation}
    \label{eq:ssgg_tree}
    \begin{split}
        \mathcal{A}_{\rm EH}^{(0)} (1^\phi, 2^\phi, 3^{--}, 4^{++}) \ & = \ - 
        \left({\kappa\over 2}\right)^2 
        {\langle 3 | 1 | 4]^4\over s^2} \Big[ {i\over t-m^2} + {i\over u - m^2} \Big]\, , 
        \\
        \mathcal{A}_{\rm EH}^{(0)} (1^\phi, 2^\phi, 3^{++}, 4^{++}) & = \ - 
        \left({\kappa\over 2}\right)^2 
        m^4 {[34]^2\over \langle 34\rangle^2} \Big[ {i\over t-m^2} + {i\over u - m^2} \Big]\, , 
    \end{split}
\end{equation} 
The computation of the four-point one-loop amplitude without helicity flip in the eikonal approximation \eqref{bendlim}  was performed in 
 \cite{Chi:2019owc}, with the result
\begin{align}
    \label{Chi}
    \begin{split}
        \mathcal{A}_{\rm EH}^{(1)}  (1^\phi, 2^\phi, 3^{--}, 4^{++})  \simeq \cN_h \left( \frac{\kappa}{2}\right)^4 \bigg[ & (2 m \omega)^4 \big( I_4(s, t; m) + I_4 (s, u; m)\big) - 15 ( m^2 \omega)^2 I_3(s; m)
        \\ & +(4 m \omega)^2 s I_3(s)  - {29\over 2} (m \omega)^2 I_2 (s) \bigg]\ , 
    \end{split}
\end{align}
where 
\begin{align}
\label{enneacca}
\cN_h := \left( \frac{\langle 3| 2| 4]}{2 m \omega}\right)^4
\ 
\end{align}
is a pure phase, with $\cN_h \to 1$  in the eikonal approximation. 
We have also computed the new amplitude with helicity flip in the same approximation, with the result 
\begin{align}
    \label{EHflip}
    \mathcal{A}_{\rm EH}^{(1)} (1^\phi, 2^\phi, 3^{++}, 4^{++}) & \simeq \ \left( \frac{\kappa}{2} \right)^4 \, \frac{\sqr{3}{4}^2}{\agl{3}{4}^2} (m^2 s)^2  \Big[ I_4(s, t;m) + I_4 (s, u;m)\Big] \ . 
\end{align}

\subsection{Four-point scalar/graviton scattering in EH + \texorpdfstring{$R^3$}{R3}}
\label{R3amplitudes}

We now consider the amplitudes with addition of the $R^3$ interaction in \eqref{action}: the helicity-preserving amplitude at tree-level is vanishing
\begin{equation}
\label{uuuu}
    \mathcal{A}_{R^3}^{(0)} (1^\phi, 2^\phi, 3^{--}, 4^{++}) = 0\ ,
\end{equation}
while the helicity-flip amplitude is  \cite{Brandhuber:2019qpg}
\begin{equation}
\label{3.6}
    \mathcal{A}_{R^3}^{(0)} (1^\phi, 2^\phi, 3^{++}, 4^{++}) = i \, 
    \left({\kappa\over 2}\right)^2 \left({\alpha^\prime\over 4}\right)^2 
     [34]^4\frac{(t-m^2)\, (u-m^2)}{s} \ .
\end{equation}
At one loop,  the result of \cite{Brandhuber:2019qpg}  for  the no-flip amplitude gives: 
\begin{equation}
    \begin{split}
    \label{gravitonbending}
        \mathcal{A}^{(1)}_{R^3}  (1^\phi, 2^\phi, 3^{--}, 4^{++}) \simeq \left({\kappa\over 2}\right)^4 \left({\alpha^\prime\over 4}\right)^2 \, \cN_h \, \Big[& (m s)^4 \big(I_4 (s, t;m) + I_4 (s, u;m) \big)  + (m^2 s\, \omega )^2  I_3 (s;m) \\ & + \frac{3}{2} (m s\, \omega )^2    I_2 (s) \Big]  \, , 
    \end{split}
\end{equation}
where $\cN_h$ is defined in \eqref{enneacca}. The one-loop amplitude with helicity flip requires a new computation and the result in the eikonal approximation is
\begin{equation}
\label{gravitonbending2}
    \begin{split}
        \mathcal{A}^{(1)}_{R^3}  (1^\phi, 2^\phi, 3^{++}, 4^{++}) \simeq & \left({\kappa\over 2}\right)^4 \left({\alpha^\prime\over 4}\right)^2 \sqr{3}{4}^4 \,  \bigg[  (2 m \omega )^4  \big(I_4 (s, t;m) + I_4 (s, u;m) \big) \\[.2em]
        & -13 ( m^2 \omega)^2  I_3 (s;m) +16 (m \omega)^2 s \, I_3(s) + \dfrac{153}{10} (m \omega)^2 I_2 (s) \bigg]
        \ .
    \end{split}
\end{equation}

\subsection{Four-point scalar/graviton scattering in EH + \texorpdfstring{$R^4$}{R4}}
\label{sec:R4ampll}
In this section we consider the addition of  an   $R^4$ interaction to the  EH action. Such interaction affects the two-scalar two-graviton amplitude at one loop  and thus contributes to graviton deflection and time delay at order $G^2$. In order to build this amplitude using the unitarity-based method we first need to find out the expression for the four-graviton tree-level amplitudes in the $R^4$ theory. We do so here starting from the Lagrangian in \eqref{L8} in order to make contact with the notation of  \cite{Endlich:2017tqa}; in Appendix \ref{app:D} we present an alternative derivation only relying on 
little-group considerations and dimensional analysis, which does not require writing down any Lagrangian.

Deriving the four-graviton amplitudes from \eqref{L8} is straightforward -- 
we simply have to replace the four Riemann tensors in each term by their linearised form corresponding to the four
on-shell gravitons.
For particle $i$ the well-known expression in momentum space is
\begin{equation}
    R(i)_{ \mu \nu \rho \sigma}\, =\, \frac{1}{2}\, F(i)_{ \mu \nu}\, F(i)_{ \rho \sigma}
\end{equation}
where
\begin{equation}
    F(i)_{\mu \nu} \,=\, p_{i \mu}\, \varepsilon_{i \nu} - p_{i \nu}\, \varepsilon_{i \mu}\ .
\end{equation}
Since we are interested in helicity amplitudes, we choose the field strengths $F(i)$ to be selfdual (negative helicity) or anti-selfdual (positive helicity), hence in spinor-helicity formalism their form is
\begin{equation}\label{A-SD}
F(i)_{\mathrm{SD}\, \alpha \dot{\alpha} \beta \dot{\beta}} = - \sqrt{2}\, \lambda_{i\alpha} \lambda_{i\beta} \epsilon_{\dot{\alpha} \dot{\beta}}
\ \ \ \mathrm{and} \ \ \
F(i)_{\mathrm{ASD}\, \alpha \dot{\alpha} \beta \dot{\beta}} = - \sqrt{2}\, \tilde{\lambda}_{i \dot{\alpha}} \tilde{\lambda}_{i \dot{\beta}} \epsilon_{\alpha \beta} \ .
\end{equation}

The building blocks in \eqref{eq:Cs} are bilinear in Riemann tensors, and take the form
\begin{equation}
    \mathcal{C} \, \simeq \left(F(i)_{\rm (A)SD} \cdot F(j)_{\rm (A)SD}\right)^2 \ ,
\end{equation}
and
\begin{equation}
    \widetilde{\mathcal{C}} \, \simeq \left(F(i)_{\rm (A)SD} \cdot F(j)_{\rm (A)SD}\right)
    \left(F(i)_{\rm (A)SD} \cdot \frac{1}{i}\ast F(j)_{\rm (A)SD}\right) \ ,
\end{equation}
where $\cdot$ denotes Lorentz contractions and $\ast$ denotes the usual Hodge dual which acts on the (anti-)selfdual field strengths as
$\ast F_{\rm SD} = F_{\rm SD}$ and $\ast F_{\rm ASD} = -F_{\rm ASD}$. Furthermore, given
the form \eqref{A-SD} these expressions are only non-vanishing if both particles $i$ and $j$ have
the same helicity. In summary, if both gravitons have negative helicity (SD field strength)  we have
\begin{equation}
\mathcal{C} = i \, \widetilde{\mathcal{C}} = \frac{1}{2}\, \agl{i}{j}^4\ ,
\end{equation}
while if both gravitons have positive helicity (ASD field strength) we have
\begin{equation}
\mathcal{C} = - i\,  \widetilde{\mathcal{C}}  =\frac{1}{2}\, [ij]^4 \ .
\end{equation}

With these results one easily arrives at 
\begin{align}
\begin{split}
    \cA_{R^4}^{(0)} (1^{++},2^{++},3^{++},4^{++}) &= i \beta^{+} \left( \frac{\kappa}{2} \right)^2 \left( \sqr{1}{2}^4 \sqr{3}{4}^4 + \sqr{1}{3}^4 \sqr{2}{4}^4 + \sqr{1}{4}^4 \sqr{2}{3}^4 \right) \ , \\
    \cA_{R^4}^{(0)} (1^{--},2^{--},3^{--},4^{--}) &= i \beta^{-} \left( \frac{\kappa}{2} \right)^2 \left( \agl{1}{2}^4 \agl{3}{4}^4 + \agl{1}{3}^4 \agl{2}{4}^4 + \agl{1}{4}^4 \agl{2}{3}^4 \right) \ ,\\
    \label{beetat}
    \cA_{R^4}^{(0)} (1^{++},2^{++},3^{--},4^{--}) &= i \widetilde{\beta} \left( \frac{\kappa}{2} \right)^2 \sqr{1}{2}^4 \agl{3}{4}^4
    \ ,
\end{split}
\end{align} 
with 
\begin{align}
\label{beta+}
         \beta^+ &=4\, \Big( \beta_1 +\frac{i}{2}\, \beta_2 - \beta_3\Big) \ ,\\[.2em]
         \label{beta-}
        \beta^- &= 4\,\Big( \beta_1 -\frac{i}{2}\, \beta_2 - \beta_3\Big)\ ,\\[.2em]
        \label{betat}
        \widetilde{\beta} &= 4\,\Big( \beta_1 + \beta_3 \Big) \ .
\end{align}
Note  that if we do not allow the parity-odd coupling ($\beta_2=0$), then the coefficient of the all-plus and all-minus amplitudes are the same $\beta^+=\beta^-:= \beta$.

The next step is to carry out one-loop amplitude calculations in the eikonal approximation, as done in previous sections.  The result for the  relevant amplitudes is:
\begin{align}
\begin{split}
\label{am1}
\mathcal{A}^{(1)}_{R^4}  (1^\phi, 2^\phi, 3^{--}, 4^{++})  &\simeq  -  \cN_h\, \widetilde{\beta} \left( \dfrac{\kappa}{2}\right)^4 \, s^2 \left[ \dfrac{35}{4} \, ( m \omega)^4 \, I_3 (s;m) + \dfrac{93}{8}  (m \omega^2)^2 \, I_2 (s) \right] 
\ ,
\\
%\def\arraystretch{2.5}
%\begin{array}{l}
   % \label{am2}
 \mathcal{A}^{(1)}_{R^4}  (1^\phi, 2^\phi, 3^{++}, 4^{++})  &\simeq -  \beta^{+}\,\left( \dfrac{\kappa}{2}\right)^4 \,\sqr{3}{4}^4 \left[ 
     %\left( 2\, \alpha m^4 s^2  + 
     \dfrac{3}{4} \, (m \omega)^4 
     %\right)
     \, I_3 (s;m) \right. 
%\\
 %	&%\hspace{6.5cm}
	+\left. \dfrac{55}{24}  \, (m\omega^2)^2 \, I_2 (s) \right] 
	\ ,
\\
%\def\arraystretch{2.5}
%\begin{array}{l}
   % \label{am2}
 \mathcal{A}^{(1)}_{R^4}  (1^\phi, 2^\phi, 3^{--}, 4^{--})  &\simeq -  \beta^{-}\,\left( \dfrac{\kappa}{2}\right)^4 
\langle 3 4\rangle^4 \left[ 
     %\left( 2\, \alpha m^4 s^2  + 
     \dfrac{3}{4} \, (m \omega)^4 
     %\right)
     \, I_3 (s;m) \right. 
%\\
 %	&%\hspace{6.5cm}
	+\left. \dfrac{55}{24}  \, (m\omega^2)^2 \, I_2 (s) \right] \ , 
\end{split}
\end{align}
where $\cN_h$ was introduced in \eqref{enneacca}.

\subsection{Scattering with the \texorpdfstring{$FFR$}{FFR} interaction}

The last interaction we wish to consider is the $FFR$ term  in \eqref{action}. 
From an on-shell point of view this is the simplest non-minimal modification of the
coupling of photons to gravity. As we will show below this leads to new corrections to the bending and time delay/advance of light and graviton propagation in the background of a very massive scalar particle.

This new interaction modifies the three-point two-photon/one-graviton amplitude:
\begin{equation}
    \cA^{(0)}_{\rm FFR} (1^+ , 2^+ , 3^{++}) = i \left(\frac{\kappa}{2}\right)\left(\frac{\alpha_\gamma}{4}\right) \sqr{1}{3}^2 \sqr{2}{3}^2\ ,
\end{equation}
which we will now use to construct the relevant amplitudes at tree level and one loop to compute deflection angles and time delay in the presence of this interaction. Note that this  amplitude is determined by its helicity structure and dimensional analysis up to a normalisation  which we fixed from the  Feynman rule \eqref{FeynFFR} following from  our action \eqref{action}.

\subsubsection{Relevant amplitudes for graviton deflection}

\label{sec:ragd}

\begin{figure}
        \centering
            \includegraphics[scale=1]{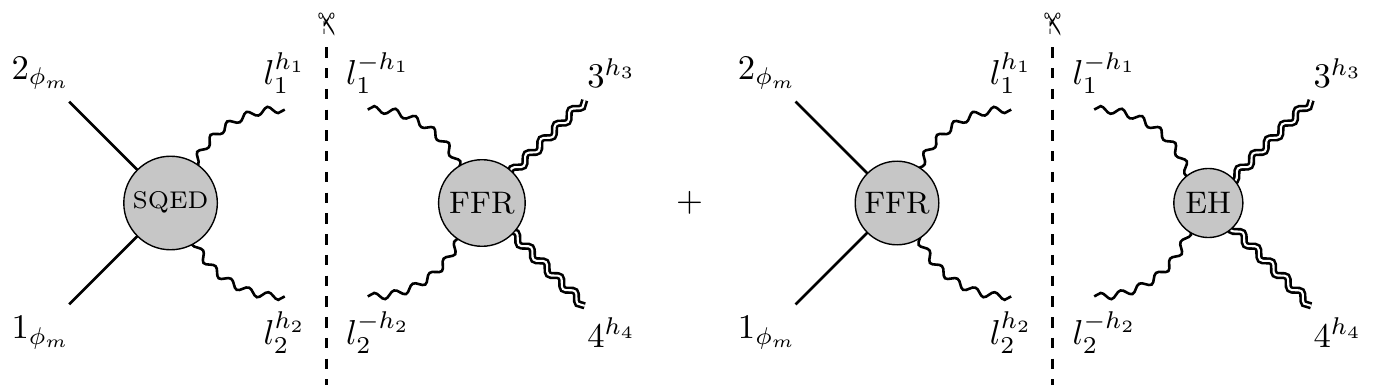}
            \caption{The two-particle cut diagrams in the $s=-\vec{q}^{\, \, 2}$-channel of the graviton deflection angle in the presence of an ${\rm FFR}$ interaction. 
           The internal lines are photons. The first diagram is proportional to $\kappa^2 \, e^2$ and is only non-vanishing for $h_1=h_2$ for the internal photons. The second diagram is proportional to $\kappa^4$, it is non-vanishing when $h_4 = - h_3$ and $h_2 = - h_1$ thus it contributes solely to the helicity-preserving configuration. Also, it only produces quantum corrections (bubble integrals) with coefficients that vanish in the case of four-dimensional external kinematics.}
            \label{fig:doublecutFFRgraviton}
        \end{figure}

Using factorisation and Feynman diagrams we have computed the four-point amplitudes relevant for graviton deflection  from a massive {\it charged} source (such as a charged black hole).
The  new ${ FFR}$ interaction  involves two photons and one graviton,  hence one cannot generate a tree-level correction to the  amplitude with two scalars and two gravitons. The  first corrections arise at one loop, from the cut diagrams in Figure \ref{fig:doublecutFFRgraviton}.

For the cut diagram on the left-hand side of the figure, we need the 
tree-level scalar QED amplitude with two photons and two massive scalars~\cite{Bjerrum-Bohr:2013bxa}
\begin{equation}
\cA^{(0)}_{\rm SQED} (1^\phi , 2^\phi , 3^{+} , 4^{+}) =  \, Q^2 \, m^2\, \frac{ \, \sqr{3}{4}^2}{s} \left( \frac{i}{t-m^2} + \frac{i}{u-m^2} \right) \> ,
\end{equation}
along with the modification  to the two-graviton/two-photon amplitudes arising from the $FFR$ coupling for both helicity configurations of the graviton: no flip,
%for the two cases where the helicity of the graviton is unchanged,
\begin{equation}\label{eq:FFRphphgrmgrp}
    \cA^{(0)}_{\rm FFR} (1^+ , 2^+ , 3^{--} , 4^{++}) = - i \left(\frac{\kappa}{2}\right)^2 \left(\frac{\alpha_\gamma}{4}\right) \sqr{1}{2}^2 \frac{\langle 3 | 1 | 4 \rbrack^4}{s t u}\ ,
\end{equation}
or flipped,
\begin{equation}\label{eq:FFRphphgrpgrp}
    \cA^{(0)}_{\rm FFR} (1^+ , 2^+ , 3^{++} , 4^{++}) = i \left(\frac{\kappa}{2}\right)^2 \left(\frac{\alpha_\gamma}{4}\right) \left(\frac{\sqr{1}{3}^2 \sqr{3}{4}^2 \sqr{4}{2}^2}{s_{1 3}} + \frac{\sqr{2}{3}^2 \sqr{3}{4}^2 \sqr{4}{1}^2}{s_{2 3}} \right)\ .
\end{equation}
Both amplitudes can be computed with on-shell techniques. Specifically, \eqref{eq:FFRphphgrmgrp} can be constructed using  BCFW  recursion relations \cite{Britto:2005fq} by shifting appropriately  the graviton momenta, while it is easy to verify \cite{Cohen:2010mi} that \eqref{eq:FFRphphgrpgrp} can be derived via an (holomorphic) all-line shift.

Note that the cut is non-vanishing only in the singlet configuration (internal photons with the same helicities). This is because the four-point amplitude with two photons and two gravitons induced by the $FFR$ interaction is non-vanishing only for same-helicity photons.

We now move to the cut diagram on the right-hand side of Figure \ref{fig:doublecutFFRgraviton}. The two-photon/two-graviton EH amplitude only exists in the configuration where the gravitons and the photons have opposite helicity (see for instance \cite{Bjerrum-Bohr:2016hpa}),
\begin{equation}\label{eq:EHphphgrgr}
\cA^{(0)}_{\rm EH} (1^{+} , 2^{-} , 3^{++} , 4^{--}) = \, -i \, \left(\dfrac{\kappa}{2}\right)^2 \, \sqr{1}{3}^2 \agl{2}{4}^2 \, \frac{ \langle 4 |1|3]^2}{stu} \> ,
\end{equation}
and thus it contributes only in the helicity-preserving process. Hence, in order to compute the cut we will only need the following 
 two-scalar/two-photon amplitude involving an $FFR$ interaction:
\begin{equation}\label{eq:FFRphiphiphph}
    \cA^{(0)}_{\rm FFR} (1^\phi , 2^\phi , 3^{-} , 4^{+}) = - i \left(\frac{\kappa}{2}\right)^2 \left(\frac{\alpha_\gamma}{4}\right) \langle 3 | 1 | 4 \rbrack^2\ .
\end{equation}
Performing the calculation, it turns out that the right-hand side of Figure \ref{fig:doublecutFFRgraviton} does not produce any non-analytic term with an $s$-channel discontinuity when external kinematics are considered to be strictly four-dimensional.

Following the above considerations, the one-loop amplitudes in the eikonal limit can be computed entirely from the LHS of Figure \ref{fig:doublecutFFRgraviton}, and are found to be 
\begin{align}\label{eq:FFR1loop}
    \begin{split}
        \cA^{(1)}_{\rm FFR} (1^\phi , 2^\phi , 3^{--} , 4^{++}) &\simeq - \cN_{h} Q^2 \left(\frac{\kappa}{2}\right)^2 \left(\frac{\alpha_\gamma}{4}\right) s \Bigg[(m s)^2 \left(I_4 (s, t;m) + I_4 (s, u;m) \right)\\[.2em]
        & \hspace{10em}+ (m \omega)^2 I_3 (s;m) + \frac{3}{4}\frac{s^3}{\omega^2} I_3(s)+ \frac{3}{2}\omega^2 I_2(s) \Bigg]\ ,\\[.2em]
        \cA^{(1)}_{\rm FFR} (1^\phi , 2^\phi , 3^{++} , 4^{++}) &= Q^2 \left(\frac{\kappa}{2}\right)^2 \left(\frac{\alpha_\gamma}{4}\right) m^2 \sqr{3}{4}^4 I_3 (s;m) \, ,
    \end{split}
\end{align}
where again $\mathcal{N}_h$ is the phase defined in \eqref{enneacca}, and  $Q$ denotes the charge of the classical source (the black hole).

\subsubsection{Relevant amplitudes for photon deflection}
\label{sec:rapd}

It is  interesting to study how this new $FFR$ interaction affects the bending and time delay/advance of light. In order to do so, we now review  the  known two-scalar/two-photon amplitudes for minimally coupled photons  \cite{Bjerrum-Bohr:2016hpa}, and present the new  corresponding  amplitudes  induced by the $FFR$ interaction, both at tree  and one-loop level.

In the following we consider processes  where the internal legs are gravitons. In the EH theory, for the two-photon two-scalar process, only the helicity-preserving amplitude is non vanishing%
\footnote{Indeed, one can check that in four dimensions the Feynman rule for two same-helicity (on-shell) photons  and one off-shell graviton $h$  is zero: $V^{\mu \nu} (1^{\pm}, 2^{\pm}, 3^{h}) = 0$, where $V^{\mu \nu} $ is given in \eqref{VVV}.},  both at tree level
\begin{equation}\label{eq:EHphotontree}
    \cA^{(0)}_{\mathrm{EH}} (1^\phi , 2^\phi , 3^{-} , 4^{+}) = i \left(\frac{\kappa}{2}\right)^2 \frac{\langle 3 | 1 | 4 \rbrack^2}{s}\ ,
\end{equation}
and at one loop  \cite{Bjerrum-Bohr:2016hpa},
\begin{equation}
    \begin{split}
    \label{fff}
        \cA^{(1)}_{\mathrm{EH}} (1^\phi , 2^\phi , 3^{-} , 4^{+}) \simeq - \cN_\gamma \left(\frac{\kappa}{2}\right)^4 &\Bigg[(2 m \omega)^4 \left(I_4 (s, t;m) + I_4 (s, u;m)\right) -15(m^2 \omega)^2 I_3 (s;m) \\
        &  + 3s(2m\omega)^2 \, I_3(s) - \frac{161}{30} (m\omega)^2 \, I_2(s) \Bigg]\ ,
    \end{split}
\end{equation}
where the phase factor $\cN_\gamma $ is
\begin{equation}
    \cN_\gamma = \left(\frac{\langle 3 | 1 | 4 \rbrack}{2 m \omega} \right)^2 \simeq -1\ .
\end{equation}
%in the eikonal approximation, and as usual \eqref{fff} is also written in the same approximation.

We now discuss the corrections to the two-scalar two-photon amplitudes arising from one insertion of the $FFR$ interaction.  
 These come  from a single graviton exchange between a minimally coupled scalar and the  $FFR$ three-point vertex.
At tree level, only the helicity-flip amplitude 
\begin{equation}
\label{uuuuu}
    \cA^{(0)}_{\rm FFR} (1^\phi , 2^\phi , 3^{+} , 4^{+}) = - i \left(\frac{\kappa}{2}\right)^2 \left(\frac{\alpha_\gamma}{4}\right) \sqr{3}{4}^2 \left[\frac{\left(t-m^2\right)\left(u-m^2\right)}{s} + m^2\right]\ ,
\end{equation}
contributes in the eikonal approximation, while the no-flip amplitude, already quoted in  \eqref{eq:FFRphiphiphph},  is a contact term  that  is subleading in the eikonal limit (it does not have a pole in $s=-|\vec{q}\, |^2$).

\begin{figure}
        \centering
            \includegraphics[scale=1]{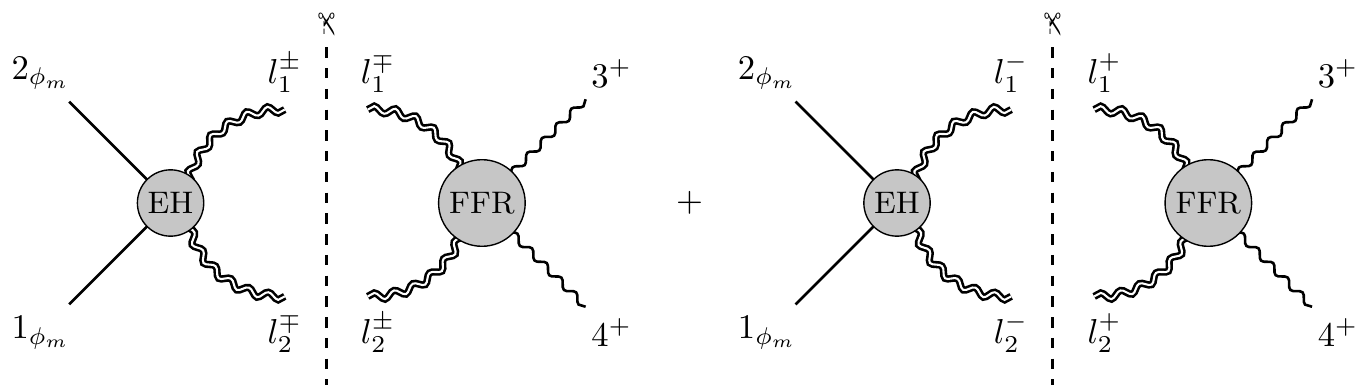}
            \caption{The two-particle cut diagrams in the $s=-|\vec{q}\, |^{\, \, 2}$-channel contributing to  photon deflection to first order in the   $FFR$ interaction. We only show  the helicity-flip configuration since the helicity-preserving cuts vanish. 
            The cut diagram on the RHS of the figure only contributes terms which are subleading in the eikonal limit.}
            \label{fig:doublecutFFRphoton}
        \end{figure}

Moving to one loop, the relevant two-particle cuts for  the $(++)$ configuration are  shown in Figure~\ref{fig:doublecutFFRphoton}. 
 We find that the  amplitude with photons in the $(++)$ helicity configuration in the eikonal approximation is 
\begin{equation}
\label{3.29}
    \begin{split}
        \cA^{(1)}_{\rm FFR} (1^\phi , 2^\phi , 3^{+} , 4^{+}) \simeq - \left(\frac{\kappa}{2}\right)^4 \left(\frac{\alpha_\gamma}{4}\right) \sqr{3}{4}^2 &\Big[(2 m \omega)^4 \left(I_4 (s, t;m) + I_4 (s, u;m) \right)\\
        & -15 (m^2 \omega)^2 I_3 (s;m) + 3\, s \, (2m\omega)^2 \, I_3(s)  \\
        & +\frac{3}{10}(m\omega)^2 \, I_2(s)\Big]\ ,
    \end{split}
\end{equation}
while the amplitude with photons in  the $(+-)$ helicity configuration  vanishes:
\begin{equation}
\label{na}
    \cA^{(1)}_{\rm FFR} (1^\phi , 2^\phi , 3^{-} , 4^{+}) = 0
    \ .
\end{equation}

\section{Eikonal phase matrix, deflection angle and time delay}
\label{sec:4}

In the previous section we have derived the relevant tree and one-loop amplitudes which we will  now use to extract the deflection angle and time delay up to 2PM order (or $\cO({G^2})$) generated by the addition of the various couplings in \eqref{action}. The key quantity  is the eikonal phase matrix $\delta$, to be introduced below, of which we will compute the leading, $\delta_0$, and subleading contributions, $\delta_1$.  
As an important consistency check we will confirm that  the leading-energy contribution of the one-loop amplitudes captures the required  exponentiation of the leading-order eikonal phase matrix $\delta_0$. 

In the following we  focus on  the classical contribution to $\delta$.
We stress that for the cases we consider, $\delta$ will be a $2\times2$ matrix: the  diagonal entries correspond to the two amplitudes $\cA(1^\phi , 2^\phi , 3^{h_1} , 4^{h_2})$
 where the helicity of the massless particle is not flipped  (which in our all-outgoing convention corresponds to $h_1=-h_2$), while the off-diagonal ones correspond to the two helicity-flip processes (with $h_1=h_2$).

As a final comment, we note that the combined effect of the interactions in \eqref{action} is simply the sum of the contributions of each interaction treated independently; hence we will study them separately, and begin our discussion by reviewing the computation in EH gravity.  

\subsection{Graviton deflection angle and time delay in Einstein-Hilbert gravity}

\subsubsection{Leading eikonal}

The relevant tree-level amplitudes in EH gravity are  given in \eqref{eq:ssgg_tree}. In the eikonal 
 approximation \eqref{bendlim} they become
\begin{equation}
    \label{eq:ssggtree}
    \begin{split}
        \mathcal{A}_{\rm EH}^{(0)} (1^\phi, 2^\phi, 3^{--}, 4^{++}) \ & 
        %= \ - \left({\kappa\over 2}\right)^2 {\langle 3 | 1 | 4]^4\over s^2} \Big[ {i\over t-m^2} + {i\over u - m^2} \Big]\to
        \simeq i \left({\kappa\over 2}\right)^2 {( 2 m \omega)^2 \over \vec{q}^{\, \, 2}} \, , \\[.2em]
        \mathcal{A}_{\rm EH}^{(0)} (1^\phi, 2^\phi, 3^{++}, 4^{++}) & 
        %= \ - \left({\kappa\over 2}\right)^2 m^4 {[34]^2\over \langle 34\rangle^2} \Big[ {i\over t-m^2} + {i\over u - m^2} \Big]\to 
        \simeq i \left({\kappa\over 2}\right)^2 {m^2 \over ( 2 \omega)^2} { q^4 \over \vec{q}^{\, \, 2}} \simeq 0 \ , 
    \end{split}
\end{equation} 
where the second amplitude is subleading compared to the first.

The  amplitudes in impact parameter space are obtained from those in momentum space using  \eqref{amptraimp}. To compute them, we will use repeatedly the result
\begin{align}
f(p, d):= \int\!{d^dq\over (2 \pi)^d} \, e^{i\vec{q} \cdot \vec{b}}  \, |\vec{q}\, |^{p}\ = \ 
{2^p \pi^{-d/2} 
\Gamma\left( 
{{d+p}\over 2}\right)\over 
\Gamma \left( - {p\over 2} \right)
} 
{1\over b^{\, d+p}} 
\ , 
\end{align}
where $b:= |\vec{b}\, |$.
We then have 
\begin{equation}
    \begin{split}
        \left.\widetilde{\mathcal{A} }_{\rm EH}^{(0)} (1^\phi, 2^\phi, 3^{--}, 4^{++})\right|_{\omega}  & 
        %= \ i  \left({\kappa\over 2}\right)^2 (m \omega) f(-2,D-2) 
        = \ i  \left({\kappa\over 2}\right)^2 {m \omega \over 4 \pi^{{D-2}\over 2}} \Gamma\left( {D\over 2} - 2 \right) {1\over b^{\, D-4}}\, , \\[.2em]
        \left.\widetilde{\mathcal{A} }_{\rm EH}^{(0)} (1^\phi, 2^\phi, 3^{++}, 4^{++})\right|_{\omega}  & = \ 0 \ , 
    \end{split}
\end{equation} 
therefore the leading eikonal phase matrix is
\begin{equation}
    \begin{split}
\label{leadingtreeEH}
\mathcal{\delta}_{0, {\rm EH}} \ = \  \left({\kappa\over 2}\right)^2 ( m \omega) f (-2, D-2) \mathbf{\uno}_2
\ \simeq - \left({\kappa\over 2}\right)^2 \frac{m \omega}{2 \pi} \left[\frac{1}{4-D} + \log{b} \right]\mathbf{\uno}_2 + \cdots \ ,
\end{split}
\end{equation}
where we omitted terms of $\cO(D-4)$ and finite terms which do not depend on $\vec{b}$.

Next we consider the one-loop amplitudes \eqref{Chi} and \eqref{EHflip}. In order to check exponentiation \eqref{exponentiation} we only keep terms that are leading in energy in the eikonal approximation, {\it i.e.}  
$\cO(\omega^3)$ in momentum space (or $\cO(\omega^2)$ in  impact parameter space). 
These  are
\begin{equation}
    \begin{split}
        \left. \mathcal{A}_{\rm EH}^{(1)} (1^\phi, 2^\phi, 3^{--}, 4^{++}) \right|_{\omega^3}& = \  \left(\frac{\kappa}{2}\right)^4 (2 m \omega)^4 \Big[ I_4(s, t; m) + I_4 (s, u; m)\Big] \ , \\[.2em]
        \left. \mathcal{A}_{\rm EH}^{(1)} (1^\phi, 2^\phi, 3^{++}, 4^{++})\right|_{\omega^3}&  
        %= \  \frac{\kappa^4}{16} (m |\vec{q}\,|)^4 \Big[ I_4(s, t) + I_4 (s, u)\Big]
        = 0 \ ,
    \end{split}
\end{equation}
where the sum of the box integrals $I_4(s, t;m) + I_4 (s, u;m)$ was evaluated in $D$ dimensions in \cite{KoemansCollado:2019ggb}
 and  is given in \eqref{boxcomb}.
Transforming to impact parameter space, we have 
\begin{align}
\left.\widetilde{\mathcal{A}}_{\rm EH}^{(1)} (1^\phi, 2^\phi, 3^{--}, 4^{++})\right|_{\omega^2}
%& = \ -\left({\kappa\over 2} \right)^4 {(m \omega)^2\over 2 \pi} f( D-6, D-2) \\&
= -\left({\kappa\over 2} \right)^4 {(m \omega)^2} \frac{2^{D-7} \Gamma ( D-4)}{\pi^{\frac{D}{2}} (D-4) \Gamma (3 - D/2)} \, {1\over b^{\, 2D-8}} \ .
\end{align} 
As expected from \eqref{exponentiation}, we find that 
\begin{align} 
\left.\widetilde{\mathcal{A}}_{\rm EH}^{(1)} (1^\phi, 2^\phi, 3^{--}, 4^{++})\right|_{\omega^2} \ = \  \frac{1}{2}\left[\left.\widetilde{\mathcal{A}}_{\rm EH}^{(0)} (1^\phi, 2^\phi, 3^{--}, 4^{++})\right|_{\omega}\right]^2 + \mathcal{O} (D-4)
\ . 
\end{align}

\subsubsection{Subleading eikonal}
\label{sec:subeikon}
In momentum space, the subleading contribution to the eikonal phase matrix is extracted from the $\cO(\omega^2)$ contribution to the amplitude in \eqref{Chi}:%
\footnote{Note that such a contribution is absent in \eqref{EHflip}.} 
\begin{align}
    \left.\mathcal{A}_{\rm EH}^{(1)} (1^\phi, 2^\phi, 3^{--}, 4^{++})\right|_{\omega^2} & = \  \left({  \kappa\over 2} \right)^4 \big( - 15 \, m^4 \, \omega^2\big) \, I_3 (s; m)  \, ,
\end{align}
where $I_3 (s; m) $ is given in \eqref{I3}, 
and as usual $s=-|\vec{q}\, |^2$. 
In the following we focus on the first term on the right-hand side of \eqref{I3}, since  the log term only contributes quantum corrections. Using 
\begin{align}
    \int\!{d^{D-2}q\over (2\pi)^{D-2}}\, e^{i\vec{q} \cdot \vec{b}}  \, |\vec{q}\, |^{-1} & =  {1\over 2 \pi} {1\over b} + \cO (D-4)
    \ , 
\end{align}
we obtain the subleading part of the amplitude in impact parameter space: 
\begin{align}
\left.\widetilde{\mathcal{A}}_{\rm EH}^{(1)} (1^\phi, 2^\phi, 3^{--}, 4^{++})\right|_{\omega} & = \ i \left({  \kappa\over 2} \right)^4 \frac{15}{256\pi} \, \frac{m^2 \omega}{b}
\ ,
\end{align}
and finally, using \eqref{delta1},  $\delta_1$:
\begin{align}\label{eq:EHgravitonsubleading}
    \delta_{1,{\rm EH}} = \left({  \kappa\over 2} \right)^4 \frac{15}{256\pi} \, {m^2 \omega\over b}\,  \bf{\uno}_2 \ .
\end{align}
The  eikonal phase matrix up to one loop in EH  is then given by 
\begin{align}
\label{Oscar}
\delta_{\rm EH} =  \delta_{0,{\rm EH}} + \delta_{1,{\rm EH}} + \cdots   =  - \left({\kappa\over 2}\right)^2 \frac{m \omega}{2 \pi} \left[\frac{1}{4-D} + \log{b} \  -  \left({  \kappa\over 2} \right)^2 \frac{15}{256\pi} \, {m\over b}\,  \right]\bf{\uno}_2  + \cdots  
\end{align}
Note that this matrix is proportional to the identity, since the polarisation of the 
gravitons scattered by the classical source is unchanged. 
The deflection angle can now be extracted using \eqref{rebend}.  While the eigenvalues of $\delta$ are divergent in $D=4$, the corresponding deflection angle is  finite: 
\begin{equation}
\label{thetaEH}
  %  \begin{split}
        \theta_{\rm EH}  \, = \, - {1\over 2 \pi} \left({\kappa\over 2}\right)^2 \, \frac{m}{b}\left[1 + \left({\kappa\over 2}\right)^2 \frac{15}{128}\, \frac{m}{b} \right]
        =\, - \frac{4\, G\, m}{b} \left(1 + G\frac{15\pi}{16} \frac{m}{b} \right)\  . 
%    \end{split} 
\end{equation}
This result agrees with the derivation of \cite{Chi:2019owc}, and as expected matches the  photon deflection angle 
\cite{Bjerrum-Bohr:2014zsa, Bjerrum-Bohr:2016hpa}, first computed by Einstein.%
\footnote{Initially up to a factor of two \cite{Einstein:1911vc}.}

Another quantity of interest which can be extracted from the eigenvalues of the eikonal matrix is the time delay. Using \eqref{eq:timedelay} applied to the  leading eikonal phase \eqref{leadingtreeEH}, we get

\begin{equation} \label{eq:EHtimedelaypole}
t_{\rm EH}=- \left({\kappa\over 2}\right)^2 \frac{m}{2 \pi}\left( \frac{1}{4-D} + \log{b}  \right) \> .
\end{equation}
As is well known, in order to define the time delay  in four dimensions we need to take the difference of two time delays as measured by an observer at $b$ and one at a much larger distance  $b_0\gg b$ \cite{Camanho:2014apa}. 
Doing so the  pole in \eqref{eq:EHtimedelaypole} drops out, and neglecting  power-suppressed terms in $b_0$ one gets
\begin{equation}
\label{tEH}
 t_{\rm EH}=\left({\kappa\over 2}\right)^2 \frac{m}{2 \pi}\log{\frac{b_0}{b}}  = {4\, G\, m} \log{\frac{b_0}{b}}  \> ,
\end{equation}
in agreement with  \cite{Ciafaloni:2014esa}. 
Including now also  the contribution from $\delta_1$, we arrive at the result
\begin{equation}
\label{zzzL}
 t_{\rm EH}= \left({\kappa\over 2}\right)^2 \frac{m}{2 \pi}\left[  \log{\frac{b_0}{b}} +\left({  \kappa\over 2} \right)^2 \frac{15}{128} \, {m \over b}\, \right]  =  {4\, G\, m}\left[  \log{\frac{b_0}{b}} +G\frac{15\pi}{16} \frac{m}{b} \right]
\> .
\end{equation}
In the next sections we compute the corrections $\Delta\theta_X  $ and $\Delta t_X $ to the deflection angle \eqref{thetaEH} and time delay \eqref{tEH} in EH due to the inclusion of  an interactions $X$ in \eqref{action}. The complete 
deflection angle and time delay will then be $\theta_{\rm EH} + \Delta \theta_X $ and $t_{\rm EH} + \Delta t_X $.

\subsection{Graviton deflection angle and time delay in \texorpdfstring{${\rm EH} + R^3$}{EH + R3}}

\subsubsection{Leading eikonal}
The relevant new amplitudes are obtained by evaluating \eqref{uuuu} and \eqref{3.6} in the eikonal limit \eqref{bendlim}, with the result
\begin{equation}
    \begin{split}
        \mathcal{A}_{R^3}^{(0)} (1^\phi, 2^\phi, 3^{--}, 4^{++}) \ & = 0\ , 
\\
\mathcal{A}_{R^3}^{(0)} (1^\phi, 2^\phi, 3^{++}, 4^{++}) & 
%= \ \left({\kappa\over 2}\right)^2 \left({\alpha^\prime\over 4}\right)^2 {4 i \over s} [34]^4(p_1\cdot p_3) (p_2 \cdot p_3) \to 
\simeq i \left({\kappa\over 2}\right)^2 \left({\alpha^\prime\over 4}\right)^2\, (2 m \omega)^2 \, {q^4 \over \vec{q}^{\, \, 2}} \ , 
    \end{split}
\end{equation} 
where from \eqref{spainors} we have $[34]^4 = q^4$. In order to transform to impact parameter space we rewrite \begin{equation}
    \vec{b}\cdot \vec{q} \ = \ \mathscr{b} \bar{q} + \bar{\mathscr{b}} q \ , 
\end{equation}
with $\mathscr{b}:= (b_1 + i b_2)/2$, and $\bar{\mathscr{b}} := (b_1 - i b_2)/2$ (and we recall our previous definitions $q=q_1 + i q_2$, $\bar{q} = q_1 - i q_2$), from which $\mathscr{b} \, \bar{\mathscr{b}} = {b^2}/{4}$. Then in $\vec{b}\,$-space we have 
\begin{equation}
    \begin{split}
        \left. \widetilde{\mathcal{A}}_{R^3}^{(0)} (1^\phi, 2^\phi, 3^{++}, 4^{++})\right|_{\omega} & = \ i \left({\kappa\over 2}\right)^2 \left({\alpha^\prime\over 4}\right)^2 ( m \omega) \left( {\partial \over \partial \bar{\mathscr{b}}}\right)^4 f(-2, D-2)\\ 
        & = \ i \left({\kappa\over 2}\right)^2 \left({\alpha^\prime\over 4}\right)^2 { ( m \omega) \over\bar{\mathscr{b}}^4} \ \xi  \ f (-2, D-2) \ ,
    \end{split}
\end{equation} 
where  
\begin{align}
\label{xi}
\xi := \Big({D\over 2} -2\Big)\Big({D\over 2} -1\Big)\Big({D\over 2} \Big)\Big({D\over 2} +1\Big)
\ .
\end{align}
Hence the leading  eikonal phase matrix  $ \delta_0$, including the first contribution from the  $R^3$ interaction, has the form
\begin{equation}
\label{leadingtreetotal}
    \delta_0 = \delta_{0, {\rm EH}} + \delta_{0, R^3}\ , 
\end{equation}
where $\delta_{0, {\rm EH}}$ is given in \eqref{leadingtreeEH}, and 
\begin{align}
\label{leadingtree}
\mathcal{\delta}_{0, R^3} \ = \  \left({\kappa\over 2}\right)^2 \left({\alpha^\prime\over 4}\right)^2 ( m \omega) \Big[\xi f (-2, D-2)\Big]  \begin{pmatrix}0 &  \bar{\mathscr{b}}^{\,-4}\\
\mathscr{b}^{- 4} & 0
 \end{pmatrix}
 \ , 
\end{align}
where  we have used \eqref{delta0}.

Moving on to one loop, 
 from \eqref{gravitonbending} and \eqref{gravitonbending2} 
we obtain 
\begin{align}
\begin{split} 
\left.\mathcal{A}_{R^3}^{(1)} (1^\phi, 2^\phi, 3^{--}, 4^{++})\right|_{\omega^3}& = 0
\ , \\
\left.\mathcal{A}_{R^3}^{(1)} (1^\phi, 2^\phi, 3^{++}, 4^{++})\right|_{\omega^3}&  = \  \Big({  \kappa\over 2} \Big)^4 \Big({  \alpha^\prime \over 4} \Big)^2 \,  [34]^4 (2 m \omega )^4 \Big[ I_4(s, t) + I_4 (s, u)\Big] \ .
\end{split} 
\end{align}
Transforming to impact parameter space, and using \eqref{boxcomb}, we arrive at 
\begin{align}
\begin{split}
\left.\widetilde{\mathcal{A}}_{R^3}^{(1)} (1^\phi, 2^\phi, 3^{++}, 4^{++})\right|_{\omega^2}&  = \ - \left({  \kappa\over 2} \right)^4 \left({  \alpha^\prime \over 4} \right)^2 {(m \omega)^2 \over 2 \pi} {1\over D-4} \left( {\partial \over \partial \bar{\mathscr{b}}}\right)^4 f(D-6, D-2)\\
&  = \ - \left({  \kappa\over 2} \right)^4 \left({  \alpha^\prime \over 4} \right)^2 {(m \omega)^2 \over 2 \pi\, \bar{\mathscr{b}}^4}
%  (D-3) (D-2) (D-1) 
\frac{\xi^\prime}{D-4}\ f(D-6, D-2)
\ ,
\end{split}
\end{align}
where
\begin{equation}
    \xi^\prime := (D-4)(D-3) (D-2) (D-1) \ .
\end{equation}
The leading  one-loop amplitude matrix in the eikonal approximation is then found to be  
\begin{align}
\mathcal{A}^{(1)}_{\omega^2}  = \ - \left({\kappa\over 2}\right)^4 
( m \omega)^2 {f (D-6, D-2) \over  2 \pi (D-4)}
\begin{pmatrix}1& \left(\dfrac{\alpha^\prime}{4}\right)^2  \dfrac{\xi^\prime}{\bar{\mathscr{b}}^4}\\
 \left(\dfrac{\alpha^\prime}{4}\right)^2  \dfrac{\xi^\prime}{\mathscr{b}^4} & 1 
 \end{pmatrix}
 \ .
\end{align}
One can then check the matrix relation 
\begin{align}
\label{expppp}
\mathcal{A}^{(1)}_{\omega^2} \ = \ - {1\over 2} (\mathcal{\delta}_0)^2 + \mathcal{O}(D-4)
\ , 
\end{align}
in agreement with \eqref{exponentiation}. In writing   \eqref{expppp} we have used that, 
 
\begin{align}
(\mathcal{\delta}_0 )^2\, =  \, \left({\kappa\over 2}\right)^4 ( m \omega)^2 \Big[ f (-2, D-2)\Big]^2
\begin{pmatrix}1& \left(\dfrac{\alpha^\prime}{4}\right)^2  \dfrac{2 \xi}{\bar{\mathscr{b}}^4}\\
\left(\dfrac{\alpha^\prime}{4}\right)^2  \dfrac{2 \xi}{{\mathscr{b}}^4} & 1 
 \end{pmatrix}
 \ ,
\end{align}
up to and including  $\mathcal{O} \big( (\alpha^\prime / 4)^2 \big)$.

Finally we compute the eigenvalues of the matrix $\delta_0$ in \eqref{leadingtreetotal}. Using 
\begin{equation}
\xi \, f(-2, D-2) = \frac{3}{2\pi} + \cO(D-4)\ ,
\end{equation} 
we can rewrite it as 
\begin{align}
\mathcal{\delta}_0 \ =  \  \left({\kappa\over 2}\right)^2 {m \omega\over 2 \pi }  
\begin{pmatrix} 
\label{maldazibo}
- \dfrac{1}{2 \epsilon} - \log b  & \left(\dfrac{\alpha^\prime}{4}\right)^2  \dfrac{3}{\bar{\mathscr{b}}^4}\\ \cr \left(\dfrac{\alpha^\prime}{4}\right)^2  \dfrac{3}{{\mathscr{b}}^4} & - \dfrac{1}{2 \epsilon} - \log b  
\end{pmatrix}\ , 
\end{align}
whose eigenvalues are
%hence the prefactor is 
%$ \left({\kappa/ 2}\right)^2 {m \omega/ ( 2 \pi) }   = 4 G ( m \omega)$. 
\begin{align}
\label{comeig}
\mathcal{\delta}_0^{(1,2)} = 
\left({\kappa\over 2}\right)^2 {m \omega\over 2 \pi }  
\Big[ -\frac{1}{2\epsilon} - \log b  \pm \left({\alpha^\prime\over 4}\right)^2  {48 \over b^4}\Big]\ .
\end{align}
Following identical steps to those leading from \eqref{Oscar} to \eqref{zzzL},
one obtains for the time delay at $\cO(G)$
\begin{align}
\label{timedelayMalda}
t_{{\rm EH}+R^3}= 
%\left({\kappa\over 2}\right)^2 {m \over 2 \pi }  
4\,  G\, m 
\Big[  \log {b_0\over b}   \pm \left({\alpha^\prime\over 4}\right)^2  {48 \over b^4}\Big]\ , 
\end{align}
where   $G = \kappa^2/ (32 \pi)$. 
For sufficiently small $b$ the eigenvalue with the choice of negative sign may become negative, leading to a time advance. 
We will come back to the time delay computation and add $\cO(G^2)$ corrections in Section \ref{sec:tdR3}.

\subsubsection{Comparison to the work of  \cite{Camanho:2014apa}}

The time advance due to $R^3$  terms was first discovered
in \cite{Camanho:2014apa},  from which it was argued that the only way to avoid causality violations is to embed the $R^3$ theory into an appropriate ultraviolet completion -- in other words a consistent ultraviolet completion of  gravitational theories with an $R^3$ interaction requires the addition of an infinite tower of massive particles with higher spins. Here we  wish to  briefly compare our results to theirs. 

The  authors of \cite{Camanho:2014apa} considered the interaction of a graviton with the background produced by a coherent state of massless particles,
and computed the eikonal phase in order to obtain  the Shapiro time delay. 
The coherent state simulates a large number of successive interactions of the graviton with a single
weakly-coupled particle, each instance being considered as independent and contributing with a small amount to the total phase shift. It is then observed that the presence of the $R^3$ coupling, which modifies the
three-point graviton amplitude, leads to non-degenerate eigenvalues of the eikonal phase matrix.

Concretely, it is interesting to compare  the eigenvalues   \eqref{comeig} of the leading eikonal phase matrix \eqref{leadingtreetotal}. 
Pleasingly, these eigenvalues   turn out to be  identical%
\footnote{Note that in   (3.22)  of \cite{Camanho:2014apa} the $1/\epsilon$ pole was not written explicitly. This pole does not affect either the  time delay \eqref{timedelayMalda} or the particle bending angle. Our  $1/\epsilon$ pole corresponds to the $\log L$ term in  \cite{Camanho:2014apa}, where $L$ is an infrared cutoff.}
to the eigenvalues  (3.22)  of \cite{Camanho:2014apa},  upon replacing $m\omega \to \omega^2$. This is due to the fact that we consider a different set-up, with massless gravitons moving in the background produced by massive scalar objects of mass $m$. In both cases the time advance is induced by the novel three-graviton coupling generated by the $R^3$ interaction.

\subsubsection{Subleading eikonal}
We now go back to the one-loop amplitudes \eqref{gravitonbending} and \eqref{gravitonbending2} 
and extract the triangle contributions which are the relevant terms contributing to the subleading eikonal matrix:
\begin{equation}
    \begin{split}
        \left. \mathcal{A}_{R^3}^{(1)} (1^\phi, 2^\phi, 3^{--}, 4^{++}) \right|_{\omega^2} &= \  \left({  \kappa\over 2} \right)^4 \left({  \alpha^\prime \over 4} \right)^2 \, \left| \vec{q}\, \right|^{4} m^4 \omega^2\, I_3 (s; m)  \ , \\[.2em]
        \left.\mathcal{A}_{R^3}^{(1)} (1^\phi, 2^\phi, 3^{++}, 4^{++}) \right|_{\omega^2} & = \ -13 \left({  \kappa\over 2} \right)^4 \left({  \alpha^\prime \over 4} \right)^2 \,   q^4 \, m^4 \omega^2\, I_3(s;m ) \ .
    \end{split}
\end{equation}

We can now transform to impact parameter space, using 
\begin{align}
\int\!{d^{D-2}q\over (2\pi)^{D-2}}\, e^{i\vec{q} \cdot \vec{b}}  \, |\vec{q}\, |^{3} & = {9\over 2 \pi} {1\over b^5} + \cO(D-4)\, , \\[.2em]
\left({\partial \over \partial \bar{\mathscr{b}}}\right)^4\int\!{d^{D-2}q\over (2\pi)^{D-2}}\, e^{i\vec{q} \cdot \vec{b}}  \, |\vec{q}\, |^{-1} & = {105\over 32 \pi}{1\over b}{1\over \bar{\mathscr{b}}^4} + \cO(D-4)\, .
\end{align}
 The amplitudes in impact parameter space then become 
\begin{equation}
    \begin{split}
        \left.\widetilde{\mathcal{A}}_{R^3}^{(1)} (1^\phi, 2^\phi, 3^{--}, 4^{++})\right|_{\omega}&  = \ -i \left({  \kappa\over 2} \right)^4 \left({  \alpha^\prime \over 4} \right)^2 \, \dfrac{9}{256\pi} \, \frac{m^2 \omega}{b^5} \ ,\\
        \left.\widetilde{\mathcal{A}}_{R^3}^{(1)} (1^\phi, 2^\phi, 3^{++}, 4^{++})\right|_{\omega}& = \ i \left({  \kappa\over 2} \right)^4 \left({  \alpha^\prime \over 4} \right)^2 \, \dfrac{1365}{4096\pi} \, \frac{m^2 \omega}{b} \, \dfrac{1}{\bar{\mathscr{b}}^4}
    \ .
     \end{split} 
\end{equation}
Using \eqref{delta0}, we can extract the contribution of the $R^3$ interaction to the subleading eikonal matrix $\delta_1$:
\begin{align}\label{eq:subleadingeikonalR3}
\mathcal{\delta}_{1,R^3}\ = \  \left({\kappa\over 2}\right)^4 \left({  \alpha^\prime \over 4} \right)^2 \dfrac{1}{256\pi}\, { m^2 \omega\over   b}
\begin{pmatrix}\ - \dfrac{9}{b^4} & \dfrac{1365}{16} \, \dfrac{1}{\bar{\mathscr{b}}^4} \ \\ \cr \dfrac{1365}{16} \, \dfrac{1}{\mathscr{b}^4}  & - \dfrac{9}{b^4} \end{pmatrix}
 \ .
\end{align}

\subsubsection{Deflection angle and time delay}
\label{sec:tdR3}

We can proceed similarly to the EH case. In the previous sections we showed that the $R^3$ interaction introduced off-diagonal terms, {\it i.e.} the helicity of the scattered graviton can change.

The eigenvalues of the leading and subleading eikonal matrices \eqref{leadingtree} and \eqref{eq:subleadingeikonalR3} are
\begin{align}
 \label{eq:R3leadingeik}   \delta_{0,R^3}^{(1,2)} &= \pm \left( \frac{\kappa}{2}\right)^2 \left( \frac{\alpha^\prime}{4}\right)^2 \frac{24}{\pi} \, \frac{m \omega}{b^4}\ , \\[.2em]
   \delta_{1,R^3}^{(1,2)} &= \left( \frac{\kappa}{2}\right)^4 \left( \frac{\alpha^\prime}{4}\right)^2 \frac{1}{256\pi}\, \frac{m^2 \omega}{b^5}\, \left( - 9 \pm 1365 \right)  \ . \label{eq:R3subleadingeik}
\end{align}

Next we present the correction to the graviton deflection angle, both in terms of $\kappa$ and $G$:
\begin{equation}
\label{main1}
    \begin{split}
        \Delta\theta_{R^3}^{(1,2)} & = - \frac{1}{2\pi} \left( \frac{\kappa}{2}\right)^2 \left( \frac{\alpha^\prime}{4}\right)^2\frac{m}{b} \left[ \pm  \frac{192}{b^4} + \frac{5}{128} (-9 \pm 1365) \left( \frac{\kappa}{2}\right)^2\frac{m}{b^5} \right]\\[.2em]
        & = - \frac{4\, G\, m}{b}  \left(\frac{\alpha^\prime}{4}\right)^2\left[ \pm  \frac{192}{b^4} + \frac{5\pi}{16} (-9 \pm 1365)\,   \frac{Gm}{b^5} \right] \ .
    \end{split}
\end{equation}

The deflection involving a graviton whose helicity is preserved in the scattering process has already been studied in \cite{Brandhuber:2019qpg}, instead the flipped helicity case is presented here for the first time.

Finally, for the time delay, proceeding as in Section \ref{sec:subeikon}, and applying 
\eqref{eq:timedelay} to \eqref{eq:R3leadingeik} and \eqref{eq:R3subleadingeik}, we arrive at
\begin{align}
\begin{split}
\label{main2}
\Delta t_{R^3}^{(1,2)}&=  \left( \frac{\kappa}{2}\right)^2 \left( \frac{\alpha^\prime}{4}\right)^2 \frac{m}{2\pi} \left[\pm 48 \, \frac{1}{b^4}+ \left( \frac{\kappa}{2}\right)^2 \frac{1}{128}\, \frac{m}{b^5}\, \left( - 9 \pm 1365 \right) \right]\\ 
&=4 G m \left( \frac{\alpha^\prime}{4}\right)^2  \left[\pm 48 \, \frac{1}{b^4}+   \frac{\pi}{16}\, \, \left( - 9 \pm 1365 \right) \frac{Gm}{b^5}\right]
\, .
\end{split}
\end{align}

\subsection{Graviton deflection angle and time delay in \texorpdfstring{${\rm EH} + R^4$}{EH + R4}}\label{sec:R4bending}

In this section we consider the deflection of gravitons induced by eight-derivative couplings in the Lagrangian, which we collectively denote as $R^4$. We will only consider the parity-even interactions in \eqref{L8} in order to present more compact formulae, therefore we set $\beta_2 = 0$, and hence $\beta^+=\beta^-=\beta$ in \eqref{beetat} and \eqref{am1}. 
Furthermore, since  these  interactions do not produce a three-graviton vertex, it is impossible to build any tree-level two-scalar two-graviton amplitude involving $R^4$. Thus there is no tree-level (1PM) bending associated to the new term in the Lagrangian, and one has
\begin{equation}
\label{442}
\delta_{0,R^4}= 0\ ,
\end{equation}
and the leading contribution arises at 2PM order.
Furthermore, since the $R^4$  term only produces a contact term 
four-graviton interaction, the resulting one-loop amplitudes does not contain any box integral. 
This is consistent with the absence of a tree-level contribution in \eqref{442} which, in the eikonal approximation, is expected to exponentiate, and would  result at one loop in the appearance of box integrals.
The same situation occurs for the 
 graviton deflection due to $FFR$ couplings discussed in Section \ref{sec:FFRgraviton}.

The relevant one-loop amplitudes are given in \eqref{am1}, and  from  the massive triangle contributions we extract the following results in the eikonal approximation:
\begin{equation}
\def\arraystretch{2.5}
\begin{array}{rl}
\left. \cA^{(1)}_{R^4} (1^\phi,2^\phi,3^{--},4^{++})\right|_{\omega^4} & = i \, \widetilde{\beta} \, \left(\dfrac{\kappa}{2}\right)^4 \, \dfrac{35}{128} \, m^3 \, \omega^4 \, |\vec{q}\, |^3  \> ,\\
\left. \cA^{(1)}_{R^4} (1^\phi,2^\phi,3^{++},4^{++})\right|_{\omega^4} & = i \, \beta \, \left(\dfrac{\kappa}{2}\right)^4 \, \dfrac{3}{128} \, m^3 \, \omega^4 \, \dfrac{q^4}{|\vec{q}\,|}  \> ,
\end{array}
\end{equation}
which then translate in impact parameter space into
\begin{align}
\begin{split}
\left. \widetilde{\cA}^{(1)}_{ R^4} (1^\phi,2^\phi,3^{--},4^{++})\right|_{\omega^3} & = i \, \widetilde{\beta} \, \left(\frac{\kappa}{2}\right)^4 \, \frac{315}{512} \, \frac{m^2 \omega ^3}{2\pi b^5}\> , \\
\left. \widetilde{\cA}^{(1)}_{ R^4} (1^\phi,2^\phi,3^{++},4^{++})\right|_{\omega^3} & = i \, \beta \, \left(\frac{\kappa}{2}\right)^4 \, \frac{315}{512}  \,  \frac{m^2 \omega ^3}{32 \pi b} \, \frac{1}{\bar{\mathscr{b}}^4} \> .
\end{split}
\end{align}
The subleading eikonal phase matrix resulting from the previous amplitudes is given by
\begin{equation}
    \label{eq:eikonalphaseR4}
\delta_{1,R^4}= \left(\frac{\kappa}{2}\right)^4 \frac{315}{512}\, \frac{m^2 \omega^3}{2\pi}\, \frac{1}{b}\, \mqty( \widetilde{\beta} \, \dfrac{1}{b^4} & \dfrac{\beta}{16} \, \dfrac{1}{\bar{\mathscr{b}}^4} \\ \cr
\dfrac{\beta}{16} \, \dfrac{1}{\mathscr{b}^4} & \widetilde{\beta} \, \dfrac{1}{b^4})\ ,
\end{equation}
whose eigenvalues are easily computed to be
\begin{equation}\label{eq:R4subleadingeik}
    \delta_{1,R^4}^{(1,2)} = \left(\widetilde{\beta} \pm \beta\right)\left(\frac{\kappa}{2}\right)^4 \frac{315}{512}\, \frac{m^2 \omega^3}{2\pi}\, \frac{1}{b^5}\ .
\end{equation}

Using \eqref{rebend} we can then extract the deflection angle
\begin{equation}
\label{main3}
    \Delta\theta_{R^4}^{(1,2)} = - \left(\widetilde{\beta} \pm \beta\right) \left(\frac{\kappa}{2}\right)^4 \frac{1575}{512}\, \frac{m^2 \omega^2}{2\pi}\, \frac{1}{b^6} =  -\left(\widetilde{\beta} \pm \beta\right)\, (Gm)^2\, \frac{1575 \pi}{16} \frac{\omega^2}{b^6}
        \ .
\end{equation}

Similarly to the EH and the $R^3$ interaction we can extract the  time delay arising from the $R^4$ interaction  in \eqref{action}, which in this case arises entirely from  the subleading eikonal phase. Applying \eqref{eq:timedelay} to \eqref{eq:R4subleadingeik} we find
\begin{equation}
\label{main4}
\Delta t_{R^4}^{(1,2)} = \left(\widetilde{\beta} \pm \beta\right) \left(\frac{\kappa}{2}\right)^4 \frac{945}{512}\, \frac{m^2 \omega^2}{2\pi}\, \frac{1}{b^5} \, = \left(\widetilde{\beta} \pm \beta\right)\,(Gm)^2\, \frac{945 \pi}{16} \,\frac{\omega^2}{b^5} 
\ .
\end{equation}
We can   express \eqref{main3} and \eqref{main4} in terms of the couplings introduced in \eqref{L8}, 
using \eqref{beta+}, \eqref{beta-} and \eqref{betat}. In the parity-even theory ($\beta_2=0$)  we get 
$\beta+\tilde\beta =8 \beta_1$, and $\tilde\beta - \beta =8 \beta_3$. 
In order to avoid a potential time-advance and associated causality violation,
we need to require
\begin{equation}
\beta_1 > 0  \quad \text{and} \quad \beta_3 >0 \, .
\end{equation}
Interestingly this positivity constraint is the same as derived from 
causality considerations in \cite{Gruzinov:2006ie} and general
$S$-matrix analyticity properties in \cite{Bellazzini:2015cra}.

\subsection{Graviton deflection angle and time delay in \texorpdfstring{${\rm EH} + { FFR}$}{EH + FFR}}\label{sec:FFRgraviton}

Next we focus our attention on  graviton deflection  in  EH theory with the addition of an $FFR$ coupling.
As discussed in Section \ref{sec:ragd}, at tree level there is no new two-scalar two-graviton amplitude generated by this interaction, hence 
\begin{equation}
\delta_{0,{\rm FFR}} = 0 \>.
\end{equation}
In order to compute  the subleading eikonal phase matrix, we look at the massive triangle contribution  to the one-loop amplitudes in \eqref{eq:FFR1loop},
\begin{equation}
    \begin{split}
        \left. \cA^{(1)}_{\rm FFR} (1^\phi,2^\phi,3^{--},4^{++})\right|_{\omega^2} &= - i \, Q^2 \left( \frac{\kappa}{2}\right)^2 \left( \frac{\alpha_\gamma}{4} \right) \frac{m \omega^2}{32} \left| \vec{q} \, \right|  \ ,\\[.2em]
        \left. \cA^{(1)}_{\rm FFR} (1^\phi,2^\phi,3^{++},4^{++}) \right|_{\omega^2} &= 0\ .
    \end{split}
\end{equation}
Using
\begin{equation}
\int\!{d^{D-2}q\over (2\pi)^{D-2}}\, e^{i\vec{q} \cdot \vec{b}}  \, |\vec{q}\, | = -\frac{1}{2 \pi} \, \frac{1}{b^3} +\cO(D-4)\ ,\\
\end{equation}
we obtain
\begin{equation}
\def\arraystretch{2.5}
\begin{array}{rl}
    \left. \widetilde{\cA}^{(1)}_{\rm FFR} (1^\phi,2^\phi,3^{--},4^{++}) \right|_{\omega} &=
    %-i \, e^2 \left( \dfrac{\kappa}{2}\right)^2 \left( \dfrac{\alpha_\gamma}{4} \right) \dfrac{\omega}{128} \, f(1,D-2)	\\
    				%&\simeq
    				i \, Q^2 \left( \dfrac{\kappa}{2}\right)^2 \left( \dfrac{\alpha_\gamma}{4} \right) \dfrac{\omega}{256 \pi}\, \dfrac{1}{b^{3}} \ , \\
     \left. \widetilde{\cA}^{(1)}_{\rm FFR} (1^\phi,2^\phi,3^{++},4^{++}) \right|_{\omega} &= 0 \> ,
\end{array}
\end{equation}
In this case the eikonal phase matrix is diagonal and the subleading contribution $\delta_{1,\mathrm{FFR}}$ is immediately seen to be
\begin{equation}\label{eq:FFRgravitoneik}
    \delta_{1,{\rm FFR}} = Q^2 \left( \frac{\kappa}{2}\right)^2 \left( \frac{\alpha_\gamma}{4} \right) \frac{\omega}{256 \pi}\, \frac{1}{b^{3}} \bf{\uno}_2\ . 
\end{equation}
The  new  contribution  to the graviton deflection angle due to the $FFR$ interaction  is then obtained using \eqref{rebend}:
\begin{equation}
\label{RN1}
    \begin{split}
        \Delta \theta_{\rm FFR} &= - Q^2 \left( \frac{\kappa}{2}\right)^2 \left( \frac{\alpha_\gamma}{4} \right) \frac{3}{256 \pi}\, \frac{1}{b^{4}}
        = - Q^2 \, G \, \left( \frac{\alpha_\gamma}{4} \right) \frac{3}{32}\, \frac{1}{b^{4}} \> .
    \end{split}
\end{equation}
Applying \eqref{eq:timedelay} to \eqref{eq:FFRgravitoneik} we find the additional contribution to  the time delay associated to the bending of a graviton in the $FFR$ theory:
\begin{equation}
\label{RN2}
\begin{split}
\Delta t_{\rm FFR} &= Q^2 \left( \frac{\kappa}{2}\right)^2 \left( \frac{\alpha_\gamma}{4} \right) \frac{1}{256 \pi}\, \frac{1}{b^{3}}
=
Q^2  G \left( \frac{\alpha_\gamma}{4} \right) \frac{1}{32}\, \frac{1}{b^{3}}
\ .
\end{split}
\end{equation}
The bending in this case is due to the electric charge $Q$ of the black hole, not to its mass, which does not appear in either \eqref{RN1} or \eqref{RN2}. We conclude  that in order to avoid possible causality violation
due to time advance the coefficient of the $FFR$ interaction must obey the positivity constraint
\begin{equation} 
\alpha_\gamma >0 \, .
\end{equation}

\subsection{Photon deflection angle and time delay in \texorpdfstring{${\rm EH} + {FFR}$}{EH + FFR}}

In this section we consider the photon deflection angle and the time delay/advance arising from the $FFR$ interaction. Compared to the case of graviton bending considered in the previous section, there is a non-vanishing tree-level contribution to the deflection, thus  we consider the leading and subleading eikonal cases separately.

\subsubsection{Leading eikonal}

The first contribution we consider arises from  the EH tree-level amplitude \eqref{eq:EHphotontree}, which in the eikonal approximation becomes%
\footnote{We recall from Section \ref{sec:rapd} that  $\cA^{(0)}_{\mathrm{EH}} (1^\phi, 2^\phi, 3^{+}, 4^{+}) = \cA^{(0)}_{\mathrm{EH}} (1^\phi, 2^\phi, 3^{-}, 4^{-})=0$.}
\begin{equation}
\label{zzzz}
    \cA^{(0)}_{\mathrm{EH}} (1^\phi, 2^\phi, 3^{-}, 4^{+}) \simeq i \left(\frac{\kappa}{2}\right)^2 \frac{(2m\,\omega)^2}{\vec{q}^{\ 2}} \> ,
\end{equation}
or,  upon transforming to impact parameter,
\begin{equation}\label{eq:EHphotontreeimpact}
    \widetilde{\cA}^{(0)}_{\mathrm{EH}} (1^\phi, 2^\phi, 3^{-}, 4^{+}) \simeq i \left(\frac{\kappa}{2}\right)^2 m\,\omega\, f(-2,D-2) \> .
\end{equation}
Note that \eqref{zzzz} has the same form as  the two-scalar two-graviton amplitude  in the eikonal approximation, first equation in  \eqref{eq:ssggtree}, as as consequence of  the equivalence principle.

At tree-level the helicity-preserving $FFR$ amplitude  \eqref{eq:FFRphiphiphph} is purely a contact term, while the helicity-flip amplitude is given in  \eqref{uuuuu}. The leading contribution in the  eikonal limit is then 
\begin{equation}
\def\arraystretch{2}
\begin{array}{rl}
\cA^{(0)}_{\rm FFR} (1^\phi, 2^\phi, 3^{-}, 4^{+}) &\simeq \, 0 \> , \\
\cA^{(0)}_{\rm FFR} (1^\phi, 2^\phi, 3^{+}, 4^{+}) &\simeq \, i \left(\dfrac{\kappa}{2}\right)^2 \left(\dfrac{\alpha_\gamma}{4}\right) (2m\,\omega)^2 \dfrac{q^2}{\left|\vec{q}\,\right|^{\, 2}}\ ,
\end{array}
\end{equation}
where we used $\sqr{3}{4}^2 = - q^2$. Transforming the non-vanishing helicity-flip amplitude to impact parameter space we obtain
\begin{equation}\label{eq:FFRphotontreeimp}
    \widetilde{\cA}^{(0)}_{\rm FFR} (1^\phi, 2^\phi, 3^{+}, 4^{+}) \simeq i \left(\frac{\kappa}{2}\right)^2 \left(\frac{\alpha_\gamma}{4}\right)\frac{m\,\omega}{\bar{\mathscr{b}}^2} \, \xi'' \, f(-2,D-2) \> ,
\end{equation}
where
\begin{equation}
\xi''=\left(\frac{D}{2}-2\right) \left(\frac{D}{2}-1\right) \> .
\end{equation}
Defining
\begin{equation}
    \delta_0^{\gamma} = \delta_{0, {\rm EH}}^{\gamma} + \delta_{0, {\rm FFR}}^{\gamma}\ , 
\end{equation}
we can combine \eqref{eq:EHphotontreeimpact} and \eqref{eq:FFRphotontreeimp} into a single leading eikonal phase matrix\footnote{There is no need here to separate the EH and the$FFR$contributions, since we consider only photon bending coming from this source.}
\begin{equation}
\delta_{0, {\rm FFR}}^{\gamma} = \left(\frac{\kappa}{2}\right)^2 m\,\omega \, f(-2,D-2) \, \mqty( 1 & \left(\dfrac{\alpha_\gamma}{4}\right) \dfrac{\xi''}{\bar{\mathscr{b}}^2} \\ \cr \left(\dfrac{\alpha_\gamma}{4}\right)  \dfrac{\xi''}{ \mathscr{b}^2} & 1) \> ,
\end{equation}
which, upon expanding around $D=4$, reduces to
\begin{equation}
\label{kkkk}
    \delta_{0, {\rm FFR}}^{\gamma} = - \left(\frac{\kappa}{2}\right)^2 \frac{m\,\omega}{2\pi} \, \mqty( \dfrac{1}{4-D} + \log{b} & -\left(\dfrac{\alpha_\gamma}{4}\right) \dfrac{1}{2 \bar{\mathscr{b}}^2} \\ \cr -\left(\dfrac{\alpha_\gamma}{4}\right) \dfrac{1}{2 \mathscr{b}^2} & \dfrac{1}{4-D} + \log{b})\ .
\end{equation}

Next, in order to test the expected exponentiation property of the leading eikonal phase matrix, we consider the terms of  $\cO(\omega^2)$ in the one-loop amplitudes. These are given in impact parameter space by
\begin{equation}
    \begin{split}
        \left. \widetilde{\cA}^{(1)}_{\mathrm{EH}} (1^\phi,2^\phi,3^{-},4^{+})\right|_{\omega^2} &= - \left(\dfrac{\kappa}{2}\right)^4 (m\,\omega)^2 \, \dfrac{f(D-6,D-2)}{2\pi (D-4)} \> , \\
        \left. \widetilde{\cA}^{(1)}_{\rm FFR} (1^\phi,2^\phi,3^{+},4^{+})\right|_{\omega^2} &= - \left(\dfrac{\kappa}{2}\right)^4 \left(\frac{\alpha_\gamma}{4}\right) \dfrac{(m\,\omega)^2}{\bar{\mathscr{b}}^2} \, \left(D-3\right) \dfrac{f(D-6,D-2)}{2 \pi} \>,
    \end{split}
 \end{equation}
 which are obtained from \eqref{fff} and \eqref{3.29}. 
In  matrix form, 
\begin{equation}
    \widetilde{\mathcal{A}}^{(1)}_{\omega^2} = - \left(\frac{\kappa}{2}\right)^2 \frac{(m\,\omega)^2}{2\pi} \, f(D-6,D-2) \, \mqty( \dfrac{1}{D-4} & \left(\dfrac{\alpha_\gamma}{4}\right) \dfrac{D-3}{ \bar{\mathscr{b}}^2} \\ \cr \left(\dfrac{\alpha_\gamma}{4}\right) \dfrac{D-3}{\mathscr{b}^2} & \dfrac{1}{D-4} )\ .
\end{equation}
Expanding around  $D=4$  we find that  $\tilde{\mathcal{A}}^{(1)}_{\omega^2}$ satisfies the matrix equation 
\begin{equation}
   \widetilde{\mathcal{A}}^{(1)}_{\omega^2} \ = \ - {1\over 2} (\mathcal{\delta}_0)^2 + \mathcal{O}(D-4)\ , 
\end{equation}
as expected. 

\subsubsection{Subleading eikonal}

Next we consider the subleading eikonal phase. The only non-vanishing EH contribution comes from the one-loop massive triangles in the helicity-preserving amplitude  \eqref{fff}, and reads
\begin{equation}
    \left. \widetilde{\cA}^{(1)}_{\mathrm{EH}} (1^\phi,2^\phi,3^{-},4^{+})\right|_{\omega} = i \left(\frac{\kappa}{2}\right)^4\frac{15}{256\pi} \, \frac{m^2 \omega}{b} \, .
\end{equation}
Just as in the case of the leading eikonal phase, the bending angle of photons in pure EH comes is the same as the graviton bending \eqref{eq:EHgravitonsubleading} thanks to the equivalence principle.

The contributions coming from the $FFR$ interaction are obtained from  \eqref{na} and \eqref{3.29}, and 
 in impact parameter space are
\begin{equation}
\def\arraystretch{2}
\begin{array}{rl}
\left. \widetilde{\cA}^{(1)}_{\rm FFR} (1^\phi,2^\phi,3^{-},4^{+})\right|_{\omega} &= \, 0 \> ,\\
\left. \widetilde{\cA}^{(1)}_{\rm FFR} (1^\phi,2^\phi,3^{+},4^{+})\right|_{\omega} &= \, i \left(\dfrac{\kappa}{2}\right)^4 \left(\dfrac{\alpha_\gamma}{4}\right) \dfrac{45}{1024\pi} \, \dfrac{m^2 \omega}{b}\, \dfrac{1}{\bar{\mathscr{b}}^2} \>.
\end{array}
\end{equation}
Combining these results into a  subleading eikonal phase matrix we get
\begin{equation}
    \delta_{1, {\rm FFR}}^{\gamma} = \left(\frac{\kappa}{2}\right)^4\frac{15}{256\pi} \, \frac{m^2 \omega}{b}\ \mqty( 1 & \left(\dfrac{\alpha_\gamma}{4}\right) \dfrac{3}{4\, \bar{\mathscr{b}}^2} \\ \cr \left(\dfrac{\alpha_\gamma}{4}\right) \dfrac{3}{4\, \mathscr{b}^2} & 1) \> .
\end{equation}

\subsubsection{Deflection angle and time delay}

Having computed   the eikonal phase matrix   at leading and subleading order,  we  can now  extract the  light bending  angle and time advance/delay. First we compute the eigenvalues of the leading eikonal phase matrix \eqref{kkkk}:
\begin{equation}\label{eq:FFRphotoneik}
    \delta_{0, {\rm FFR}}^{\gamma \,(1,2)} = - \left(\frac{\kappa}{2}\right)^2 \frac{m\,\omega}{2\pi} \left[\left(\frac{1}{4-D} + \log b \right) \mp \left(\frac{\alpha_\gamma}{4}\right) \frac{2}{b^2} \right] \>,
\end{equation}
which match qualitatively the result of photon deflection in a shockwave background (see \cite{Hollowood:2015elj}, and \cite{Goon:2016une} for related work), while  at subleading order we have
\begin{equation}\label{eq:FFRphotonsubeik}
    \delta_{0, {\rm FFR}}^{\gamma \,(1,2)} = \left(\frac{\kappa}{2}\right)^4\frac{15}{256\pi} \, \frac{m^2 \omega}{b} \left[1 \pm \left(\frac{\alpha_\gamma}{4}\right) \frac{3}{b^2}\right] \>.
\end{equation}
Using once again \eqref{rebend}, we find the light bending angle up to $\cO(G^2)$:
\begin{equation}
\label{main7}
    \begin{split}
      \Delta  \theta^{\gamma\, (1,2)}_{{\rm FFR}} &= - \left(\frac{\kappa}{2}\right)^2 \frac{1}{2\pi} \frac{m}{b} \left\{ 1 \pm \left( \frac{\alpha_\gamma}{4}\right) \frac{4}{b^2} + \left(\frac{\kappa}{2}\right)^2 \frac{15}{128}\,\frac{m}{b} \left[1\pm \left(\dfrac{\alpha_{\gamma}}{4}\right) \frac{9}{b^2}\right]\right\}\\[.2em]
        &= - \frac{4\, G\, m}{b} \left\{ 1 \pm \left( \frac{\alpha_\gamma}{4}\right) \frac{4}{b^2} + \frac{15\pi}{16}\, \frac{Gm}{b} \left[1\pm \left(\dfrac{\alpha_{\gamma}}{4}\right) \frac{9}{b^2}\right]\right\}\ .
    \end{split}
\end{equation}

Finally, applying \eqref{eq:timedelay} to \eqref{eq:FFRphotoneik} and \eqref{eq:FFRphotonsubeik} we arrive at our result for the   time delay:
\begin{equation}
\label{main8}
\begin{split}
 \Delta t^{\gamma \,(1,2)}_{{\rm FFR}} &=  \left(\frac{\kappa}{2}\right)^2 \frac{m}{2\pi} \left\{ \log{\frac{b_0}{b}} \pm \left(\frac{\alpha_\gamma}{4}\right) \frac{2}{b^2} +\left(\frac{\kappa}{2}\right)^2\frac{15}{128} \, \frac{m}{b} \left[1 \pm \left(\frac{\alpha_\gamma}{4}\right) \frac{3}{b^2}\right] \right\} \\
 &= 4\,  G\, m\, \left\{ \log{\frac{b_0}{b}} \pm \left(\frac{\alpha_\gamma}{4}\right) \frac{2}{b^2} +\frac{15\pi }{16} \, \frac{Gm}{b} \left[1 \pm \left(\frac{\alpha_\gamma}{4}\right) \frac{3}{b^2}\right] \right\} \ .
\end{split}
\end{equation}
We note that the  $\cO(G\, \alpha_\gamma)$ part of our result \eqref{main7} is in precise agreement with \cite{Drummond:1979pp} while it disagrees with \cite{Berends:1975ah}.%
\footnote{The result of \cite{Berends:1975ah} for $\Delta  \theta^{\gamma}_{{\rm FFR}}$ was already identified as incorrect in \cite{Drummond:1979pp} due to an inappropriate definition of the deflection angle.}
Note that \eqref{main8} generically leads to a potential time advance and causality violation independent
of the sign of the coupling $\alpha_\gamma$. This parallels the situation for the $R^3$ interaction which
requires an appropriate UV completion to restore causality \cite{Camanho:2014apa}.

\newpage 

\section*{Acknowledgements}

We would like to thank Tim Clifton, Claudia de Rham, Lance Dixon, Gregory Korchemsky, David Kosower, Lorenzo Magnea, Rodolfo Russo, Chris White and Alexander Zhiboedov for interesting discussions and comments. We also thank the organisers of the Paris winter workshop ``The Infrared in QFT'', where our results were presented. This work  was supported by the Science and Technology Facilities Council (STFC) Consolidated Grant ST/P000754/1 \textit{``String theory, gauge theory \& duality''}, and by the European Union's Horizon 2020 research and innovation programme under the Marie Sk\l{}odowska-Curie grant agreement No.~764850 {\it ``\href{https://sagex.org}{SAGEX}''}.

%\newpage

\appendix

\section{Relevant integrals}
\label{app:A}

In this section we give the explicit expression for the integral functions appearing in our results. These expressions are expanded in $\epsilon$ up to the relevant orders, and only terms with an $s$-channel discontinuity are kept.
{\setlength{\jot}{10pt}
\begin{align}
I_2(s) &  \ \simeq \ \frac{i}{16 \pi^2} \Big[ \frac{1}{\epsilon} - \log ( - s)  \Big]\ , \\
I_3 (s) &  \simeq \ 
{i\over 16 \pi^2\, s} \Big[ \frac{1}{\epsilon^2} - \frac{\log ( -s)}{\epsilon} + \frac{1}{2} \log^2 ( -s ) \Big]\, , 
\\
I_3 (s; m) \label{I3}
&\simeq
- \frac{i}{32} \Big[ \frac{1}{m  \sqrt{ -s} } + \frac{ \log ( - {s/m^2} )}{\pi^2 m^2} \Big] \ , \\
\begin{split}
\label{boxcomb}
I_4(s, t; m) + I_4 (s, u; m) & \simeq  - {\frac{1}{8\pi}} {\frac{1}{m\omega}} {1\over D-4} {(-s)^\frac{D-6}{2} }\\
& \simeq \  -\frac{1}{16 \pi \, s \, ( m \omega)} \, \Big[  \frac{1}{\epsilon} 
\, -  \, 
\log \left(-\frac{s}{m^2}\right)\Big]\ .
\end{split}
\end{align}
}

\section{Feynman rules}
\label{app:B}

Below we list some of the Feynman rules used to obtain the new tree-level
amplitudes quoted in the paper. Note that $1_{\phi_m}$ represents a massive scalar with momentum $p_1$, $1^\alpha$ represents a photon  with momentum $p_1$, and $3^{\mu\nu}$ represents a graviton with momentum $p_3$:
\begin{equation}
    \begin{tikzpicture}[scale=22.5,baseline={([yshift=-1mm]centro.base)}]
    
        \def\x{0}
        \def\y{0}
    
        \node at (0+\x,0+\y) (centro) {};
        \node at (-1.5pt+\x,1.5pt+\y) (part2) {$2_{\phi_m}$};
        \node at (-1.5pt+\x,-1.5pt+\y) (part1) {$1_{\phi_m}$};
        \node at (2pt+\x,0+\y) (part3) {$3^{\mu \nu}$};

        \draw [thick] (centro.center) -- (part1);
        \draw [thick] (centro.center) -- (part2);
        \draw [vector] (centro.center) -- (part3);

        \node at (0.5pt+\x,0.5pt+\y) {\small ${\rm EH}$};
        
        \node at (7.6pt+\x,0pt+\y) {$= \ i\left(\dfrac{\kappa}{2} \right)\Big[ -\eta^{\mu \nu} (p_1\cdot p_2 + m^2) + p_1^\mu p_2^\nu + p_2^\nu p_1^\mu\Big]$};
    \end{tikzpicture}
\end{equation}

\begin{equation}
\label{VVV}
    \begin{tikzpicture}[scale=22.5,baseline={([yshift=-1mm]centro.base)}]
    
        \def\x{0}
        \def\y{0}
    
        \node at (0+\x,0+\y) (centro) {};
        \node at (-1.5pt+\x,1.5pt+\y) (part2) {$2^{\beta}$};
        \node at (-1.5pt+\x,-1.5pt+\y) (part1) {$1^{\alpha}$};
        \node at (2pt+\x,0+\y) (part3) {$3^{\mu \nu}$};

        \draw [vector2] (centro.center) -- (part1);
        \draw [vector2] (centro.center) -- (part2);
        \draw [vector] (centro.center) -- (part3);

        \node at (0.5pt+\x,0.5pt+\y) {\small ${\rm F}^2$};
        
        \node at (8.55pt+\x,-0.65pt+\y) {$\begin{array}{rl}
 =\, i \, \left(\dfrac{\kappa}{2} \right) \!\!\!\!& \bigg[  \dfrac{1}{2}\,\eta^{\alpha \beta}\, \eta^{\mu \nu}\, s_{1 2} - \eta^{\alpha (\mu}\, \eta^{\nu ) \beta}\, s_{1 2} -2\, \eta^{\alpha \beta}\, p_{1}^{( \mu}\, p_{2}^{\nu )}\\[1em]
  &\ +\,2\, \eta^{\alpha (\mu} \, p_{2}^{\nu )} \, p_{1}^{\beta} +2\, \eta^{\beta (\mu}\, p_{1}^{\nu )}\, p_{2}^{\alpha} -\eta^{\mu \nu}\, p_{1}^{\beta}\, p_{2}^{\alpha} \bigg]
\end{array}$};
    \end{tikzpicture}
\end{equation}

\begin{equation}
    \label{FeynFFR}
    \begin{tikzpicture}[scale=22.5,baseline={([yshift=-1mm]centro.base)}]

\def\x{0}
\def\y{0}

\node at (0+\x,0+\y) (centro) {};
\node at (-1.5pt+\x,1.5pt+\y) (part2) {$2^{\beta}$};
\node at (-1.5pt+\x,-1.5pt+\y) (part1) {$1^{\alpha}$};
\node at (2pt+\x,0+\y) (part3) {$3^{\mu \nu}$};

\draw [vector2] (centro.center) -- (part1);
\draw [vector2] (centro.center) -- (part2);
\draw [vector] (centro.center) -- (part3);

 \node at (0.5pt+\x,0.5pt+\y) {\small FFR};
 
\node at (8.2pt+\x,-0.65pt+\y) {$\begin{array}{rl}
 = \, i \left(\dfrac{\kappa}{2} \right) \left( \dfrac{\alpha_\gamma}{4}\right) \!\!\!\!& \bigg[ \eta^{\alpha (\mu}\, \eta^{\nu ) \beta}\, s_{1 3} \, s_{2 3}- 2 \eta^{\alpha ( \mu}\, p_{2}^{\nu )}\, p_{3}^{\beta}\, s_{1 3}  \\[1em]
 &\ -2 \eta^{\beta ( \mu }\, p_{1}^{\nu )}\, p_{3}^{\alpha}\, s_{2 3} +4p_{1}^{( \mu}\, p_{2}^{\nu )}\, p_{3}^{\alpha}\, p_{3}^{\beta} \bigg]
\end{array}$};

\end{tikzpicture}
\end{equation}

\section{The tree-level amplitudes}
\label{app:C}

In this appendix we collect for the reader's convenience all the tree-level amplitudes we have used in our derivations. All are consistent with the normalisations of \eqref{action}, also we assume all momenta to be outgoing.

{\setlength{\jot}{10pt}
\begin{align}
& \cA^{(0)}_{\rm SQED} (1^\phi , 2^\phi , 3^{+} , 4^{+}) =  \, e^2 \, m^2\, \frac{ \, \sqr{3}{4}^2}{s} \left( \frac{i}{t-m^2} + \frac{i}{u-m^2} \right) \> , \\
 & \cA^{(0)}_{\rm EH} (1^{+} , 2^{-} , 3^{++} , 4^{--}) = \, -i \, \left(\dfrac{\kappa}{2}\right)^2 \, \sqr{1}{3}^2 \agl{2}{4}^2 \, \frac{ \langle 4 |1|3]^2}{stu} \> , \\
  &\mathcal{A}_{\rm EH}^{(0)} (1^\phi, 2^\phi, 3^{--}, 4^{++}) \  = \ - 
        \left({\kappa\over 2}\right)^2 
        {\langle 3 | 1 | 4]^4\over s^2} \Big[ {i\over t-m^2} + {i\over u - m^2} \Big]\, , 
        \\
 & \cA^{(0)}_{\rm EH} (1^{++} , 2^{++} , 3^{--} , 4^{--}) = \, i \, \left(\dfrac{\kappa}{2}\right)^2 \, \frac{s_{1 2} s_{1 3}}{s_{1 4}}\, \frac{\agl{3}{4}^8}{\agl{1}{2}^4 \agl{2}{3}^4 \agl{3}{4}^4 \agl{4}{1}^4}\ ,\\
   &     \mathcal{A}_{\rm EH}^{(0)} (1^\phi, 2^\phi, 3^{++}, 4^{++})  = \ - 
        \left({\kappa\over 2}\right)^2 
        m^4 {[34]^2\over \langle 34\rangle^2} \Big[ {i\over t-m^2} + {i\over u - m^2} \Big]\, , \\
    &    \cA^{(0)}_{\mathrm{EH}} (1^\phi , 2^\phi , 3^{-} , 4^{+}) = i \left(\frac{\kappa}{2}\right)^2 \frac{\langle 3 | 1 | 4 \rbrack^2}{s}\ , \\
     &    \cA^{(0)}_{\mathrm{EH}} (1^\phi , 2^\phi , 3^{+} , 4^{+}) = 0\ , \\
     &A_{R^3} ( 1^{++}, 2^{++}, 3^{++}, 4^{--} ) \ = \ - i \left( \frac{\kappa}{2}\right)^2 \Big( \frac{\alpha^\prime}{4} \Big)^2
 ( \langle 4 1\rangle [1 3] \langle 3 4\rangle)^2 \frac{ [ 1 2] [2 3] [ 3 1] }{\langle 1 2\rangle \langle 2 3\rangle \langle3 1\rangle} \ , \\
      &   \mathcal{A}_{R^3}^{(0)} (1^\phi, 2^\phi, 3^{--}, 4^{++})  = 0\ , \\
       &  \mathcal{A}_{R^3}^{(0)} (1^\phi, 2^\phi, 3^{++}, 4^{++}) = i \, \left({\kappa\over 2}\right)^2 \left({\alpha^\prime\over 4}\right)^2 
         [34]^4\frac{(t-m^2)\, (u-m^2)}{s} \ , \\
       &  \cA_{R^4}^{(0)} (1^{++},2^{++},3^{++},4^{++}) = i \beta \left( \frac{\kappa}{2} \right)^2 \left( \sqr{1}{2}^4 \sqr{3}{4}^4 + \sqr{1}{3}^4 \sqr{2}{4}^4 + \sqr{1}{4}^4 \sqr{2}{3}^4 \right) \ , \\
       &  \cA_{R^4}^{(0)} (1^{++},2^{++},3^{--},4^{--}) = i \widetilde{\beta} \left( \frac{\kappa}{2} \right)^2 \sqr{1}{2}^4 \agl{3}{4}^4
    \ , \\
    & \cA_{R^4}^{(0)} (1^{\phi},2^{\phi},3^{h_3},4^{h_4}) = 0 \hspace{0.5cm}\text{with}\hspace{0.5cm} h_3,h_4 \in \{+,-,++,--\} \> , \\
   & \cA^{(0)}_{\rm FFR} (1^+ , 2^+ , 3^{++}) = i \left(\frac{\kappa}{2}\right)\left(\frac{\alpha_\gamma}{4}\right) \sqr{1}{3}^2 \sqr{2}{3}^2\ , \\
    & \cA^{(0)}_{\rm FFR} (1^+ , 2^+ , 3^{--} , 4^{++}) = - i \left(\frac{\kappa}{2}\right)^2 \left(\frac{\alpha_\gamma}{4}\right) \sqr{1}{2}^2 \frac{\langle 3 | 1 | 4 \rbrack^4}{s t u}\ , \\
    &  \cA^{(0)}_{\rm FFR} (1^+ , 2^+ , 3^{++} , 4^{++}) = i \left(\frac{\kappa}{2}\right)^2 \left(\frac{\alpha_\gamma}{4}\right) \left(\frac{\sqr{1}{3}^2 \sqr{3}{4}^2 \sqr{4}{2}^2}{s_{1 3}} + \frac{\sqr{2}{3}^2 \sqr{3}{4}^2 \sqr{4}{1}^2}{s_{2 3}} \right)\ , \\
     &  \cA^{(0)}_{\rm FFR} (1^\phi , 2^\phi , 3^{-} , 4^{+}) = - i \left(\frac{\kappa}{2}\right)^2 \left(\frac{\alpha_\gamma}{4}\right) \langle 3 | 1 | 4 \rbrack^2\ , \\
     &  \cA^{(0)}_{\rm FFR} (1^\phi , 2^\phi , 3^{+} , 4^{+}) = - i \left(\frac{\kappa}{2}\right)^2 \left(\frac{\alpha_\gamma}{4}\right) \sqr{3}{4}^2 \left[\frac{\left(t-m^2\right)\left(u-m^2\right)}{s} + m^2\right]\ .
\end{align}
}

\section{The four-graviton  amplitudes in \texorpdfstring{$R^4$}{R4}}
\label{app:D}

In this appendix we show how the most generic  four-graviton amplitude in  an $R^4$ background can be constructed just from little-group considerations and dimensional analysis, without looking at any Lagrangian. 
We begin  by noting that the  coupling constant of the four-point amplitude has two powers of $\kappa$ ($[\kappa ] = -1$) and it is proportional to the  coupling constant of the $R^4$ interaction $\beta$ ($[\beta ] = -6 $). Furthermore, the nature of the new interaction implies  that the four-point amplitude is just a contact term. Mass dimension  and  scaling under little-group transformations fix the form of the possible amplitudes completely:
\begin{align}
  \label{R4_tree_pppp}  \cA_{R^4}^{(0)} (1^{++},2^{++},3^{++},4^{++}) &= i \beta \left( \frac{\kappa}{2} \right)^2 \widetilde{\lambda}_{1}^{\, \otimes 4} \, \widetilde{\lambda}_{2}^{\, \otimes 4} \, \widetilde{\lambda}_{3}^{\, \otimes 4} \, \widetilde{\lambda}_{4}^{\, \otimes 4}\, , \\[.2em]
    \cA_{R^4}^{(0)} (1^{++},2^{++},3^{++},4^{--}) &= 0 \, , \\[.2em]
    \label{R4_tree_ppmm}
    \cA_{R^4}^{(0)} (1^{++},2^{++},3^{--},4^{--}) &= i \widetilde{\beta} \left( \frac{\kappa}{2} \right)^2 \widetilde{\lambda}_{1}^{\ \otimes 4} \widetilde{\lambda}_{2}^{\ \otimes 4} \lambda_{3}^{\ \otimes 4} \lambda_{4}^{\ \otimes 4} \ .
\end{align}
We can now introduce the convenient variables
\begin{align}
 a := \sqr{1}{2}\sqr{3}{4}\, , \qquad  b:=-\sqr{1}{3}\sqr{2}{4}\, , \qquad  c:= \sqr{1}{4}\sqr{2}{3}\ , 
 \end{align} in terms of which the all-plus amplitude can be written in such a way that  permutation invariance is manifest. By saturating the spinor indices of \eqref{R4_tree_pppp} with the Levi-Civita tensor in all  possible ways one gets four distinct combinations:
\begin{equation}
\label{fourr}
    \cA_{R^4}^{(0)} (1^{++},2^{++},3^{++},4^{++}) = i \beta \left( \frac{\kappa}{2} \right)^2
        \begin{cases}
            a^4 + b^4 + c^4\\[.1em]
            a^2\, b^2 + a^2\, c^2 + b^2\, c^2\\[.1em]
            a^3\, b + a\, b^3 + a^3\, c + a\, c^3 + b^3\, c + b\, c^3\\[.1em]
            a^2\, b\, c + a\, b^2\, c + a\, b\, c^2  \ .
        \end{cases}
\end{equation}
However, using the Schouten identity, which in terms of these variables reads
\begin{equation}
    a+b+c=0\ ,
\end{equation}
one can show that there is actually only one independent combination, which we will take to be the first of \eqref{fourr}. We will then define  the all-plus amplitude to be 
\begin{equation}
\label{beetabis}
    \cA_{R^4}^{(0)} (1^{++},2^{++},3^{++},4^{++}) = i \beta \left( \frac{\kappa}{2} \right)^2 \left( \sqr{1}{2}^4 \sqr{3}{4}^4 + \sqr{1}{3}^4 \sqr{2}{4}^4 + \sqr{1}{4}^4 \sqr{2}{3}^4 \right) \ .
\end{equation} 
In the presence of a parity-invariant theory, the  amplitude corresponding to \eqref{beetabis} with all helicities flipped is simply
obtained by replacing $[j i] \rightarrow \langle i j \rangle$, otherwise it should be considered to have an independent normalisation. 

For the  MHV amplitude  \eqref{R4_tree_ppmm} there is only one possible structure, and we define the corresponding amplitude as 
\begin{equation}
    \cA_{R^4}^{(0)} (1^{++},2^{++},3^{--},4^{--}) = i \widetilde{\beta} \left( \frac{\kappa}{2} \right)^2 \sqr{1}{2}^4 \agl{3}{4}^4
    \ .
\end{equation}
The derivation of these amplitudes from the Lagrangian  \eqref{L8} is presented in Section \ref{sec:R4ampll}.

%%%%%%%%%%%%%%%%%%%%%%%%%%%%%

\newpage

\bibliographystyle{utphys}
\bibliography{remainder}

\end{document}